\documentclass[aps,%
twocolumn,%
prc,%
floatfix,%
amsmath,%
amssymb,%
amsfonts,%
showpacs,%
superscriptaddress,%
nofootinbib,%
nobibnotes,%
eqsecnum%
]{revtex4-1}

\usepackage{graphicx}
\usepackage[pdfstartview=FitH]{hyperref}
\usepackage{mathrsfs}
\usepackage{dcolumn}

\newcommand{\Tr}{\mathop{\mathrm{Tr}}}
\newcommand{\nt}[1]{\lefteqn{#1}\slash}
\newcommand{\tabd}{\multicolumn{1}{c}{\text{---}}}

\begin{document}

\title{Gauge- and point-invariant vertices of nucleon-to-resonance interactions}

\author{G.~Vereshkov}
\email{gveresh@gmail.com}
\affiliation{%
Research Institute of Physics,
Southern Federal University,
Prospekt Stachki, 194, 344090 Rostov-na-Donu, Russia
}
\affiliation{%
Institute for Nuclear Research of the Russian Academy of Sciences,
Prospekt 60-letiya Oktyabrya, 7a, 117312 Moscow, Russia
}
\author{N.~Volchanskiy}
\email{nikolay.volchanskiy@gmail.com}
\affiliation{%
Research Institute of Physics,
Southern Federal University,
Prospekt Stachki, 194, 344090 Rostov-na-Donu, Russia
}

\begin{abstract}
We construct interactions of nucleons $N$ with higher-spin resonances $R$ invariant under point and gauge transformations of the Rarita-Schwinger field. It is found for arbitrarily high spin of a resonance that the requirement of point- and gauge-invariance uniquely determines a Lagrangian of $NR$ interactions with pions, photons, and vector mesons, which might reduce model ambiguity in effective-field calculations involving such vertices. Considering the $NR$ interactions with photons and vector mesons, the symmetry provides a classification of three $NR$ vertices in terms of their differential order. The $Q^2$ dependencies of the point and gauge invariant form factors are considered in a vector-meson--dominance model. The model is in good agreement with experimental data. In addition, we point out some empirical patterns in the $Q^2$ dependencies of the form factors: low-$Q^2$ scaling of the $N\Delta(1232)$ form factor ratios and relations between form factors for $NN(1520)$ and $NN(1680)$ transitions.
\end{abstract}



\maketitle


\section{Introduction}

Baryon-meson effective-field theory is one of the most important approaches of hadron phenomenology adopted to extract form factors from experimental data and to model $Q^2$ dependencies of the form factors (dispersion relation approach, chiral perturbation theory, etc.) \cite{bu-04, pa-07, 2012PrPNP..67....1A}. However, there are a number of barriers to describing higher-spin nucleon resonances as effective fields, with the major barrier being problems with physical and mathematical consistency of the interaction Lagrangians. Most of these problems originate from the fact that spin-$\frac{H}{2}$ baryon resonances ($H$ is equal to or greater than 3) are commonly described in terms of the reducible Rarita-Schwinger (RS) representations of the Lorentz group \cite{ra-sh}. The RS fields comprise a spin cascade of irreducible states,
\begin{align}
	J^P = \frac{H}{2}, \, \left( \frac{H}{2}-1 \right)^\pm, \, \dots \left( \frac{1}{2} \right)^\pm.
\end{align}
The lower-spin components should be eliminated by subsidiary conditions imposed on the field.

While convenient to maintain manifest Lorentz invariance of the theory, the RS fields are mathematically complicated structures possessing nontrivial intrinsic symmetries. The symmetries are inextricably linked to the constraints that are imposed on the reducible RS field to do away with the redundant degrees of freedom (DOF). This is, perhaps, the most troublesome aspect of dealing with baryon resonances given by the RS fields, because the symmetries and constraints can be broken in the presence of interactions, which consequently lead to the excitation of the lower-spin components. This may result in superluminal propagation of the resonance \cite{ve-zw, 1973singh} and make the quantization of the field impossible \cite{johnson-sudarshan, 1998pascalutsa}.

For decades many efforts have been directed towards construction of the consistent interactions of the RS fields. Most of the works seek the physical interpretation for the lower-spin components of the field \cite{1980ranada-sierra2, 2006rico-kirchbach, 2006kruglov, Haberzettl1998}. However, the most theoretically appealing solutions \cite{pa-ti, 1968peccei, 2011PhRvC..84d5201V} to the problem follow another way; these do not invoke additional DOF and rely on the interaction Lagrangians that possess the same symmetries as the free-field RS theory. In particular, Peccei \cite{1968peccei} suggested that consistent interaction Lagrangians should be invariant under point transformations of the RS field.\footnote{To avoid confusion, it should be noted that throughout this paper point invariance is taken to mean the invariance under point transformations of the RS field \emph{without shifting the free-field parameter simultaneously} (cf. Refs. \cite{1971nath-etemadi, 1989benmerrouche}). Thus, the equivalence class of the free-field Lagrangians is point invariant, while the Lagrangian itself is not for any value of the free-field parameter.} However, it was shown \cite{1971nath-etemadi, 1989benmerrouche, Haberzettl1998} that the suggestion by Peccei still does not exclude all the components of the interacting RS fields. Much later, Pascalutsa \cite{pa-ti} proposed the interaction Lagrangians that are invariant under gauge transformation of the RS fields. Such interactions do not involve lower-spin components of the RS fields, although for a specific value of off-shell parameters.

The purpose of this paper is to write Lagrangians for the interactions of higher-spin fields with nucleons that are invariant under both gauge and point transformations of the field. Thus, this work can be regarded as a synthesis of the ideas expressed by Peccei \cite{1968peccei} and Pascalutsa \cite{pa-ti}. This synthesis results in the most constrained interactions preserving the highest degree of symmetry of the free RS field. Such interactions are valid and universal in any formulation of the RS theory (in theories with \cite{1974PhRvD...9..910S} or without \cite{ra-sh, 2004pilling, *2005pilling} an \textit{a priori} tracelessness condition imposed on the field; in the second-order formalism \cite{2006napsuciale}). The point- and gauge-invariant interactions do not modify free-field covariant constraints. This property might be important, because constructing consistent nonminimal electromagnetic couplings of the RS fields without adding new DOF appears to require that the free-field $\gamma$-tracelessness condition is preserved by interactions \cite{2009PhRvD..80b5009P, 1992PhRvD..46.3529F}.

The point- and gauge-invariant interactions were studied earlier for spin-$\frac32$ and -$\frac52$ in Refs.~\cite{2009PhRvC..80e8201S, 2010PhRvC..82a5203S, 2009vereshkov, 2010arXiv1006.3929V}. In this work, we generalize the results of Refs.~\cite{2009vereshkov, 2010arXiv1006.3929V} to the case of arbitrarily high spin of the resonance. An outline of the rest of the paper is given below.

In Sec.~\ref{RS formulation} we briefly review the canonical theory of the free RS field. For the simplest case of spin-$\frac32$, we rederive the well-known facts that (1) the free-field theory is generated by a one-parameter equivalent class of Lagrangians $\mathscr{L}_\text{ff}(A)$, with the free-field parameter $A$ amounting to a sort of rotation between the lower-spin components \cite{ra-sh, moldauer-case, johnson-sudarshan}; (2) the theory is invariant under gauge transformations of the field in the massless limit \cite{ra-sh}; (3) the point transformations form the symmetry group of the equivalent class of the Lagrangians \cite{moldauer-case, johnson-sudarshan, 2004pilling, *2005pilling}.

Finally, in Sec.~\ref{RS formulation} it is shown that point and gauge invariance of the interactions result in the restrictions on the tensor-spinor source $\delta S_\text{int} /\delta \bar\Psi_{\mu_1\dots\mu_\ell}$, so that the simple subsidiary conditions on the free RS field hold in the presence of the interactions.

In Sec.~\ref{lagrangians} we proceed to explicitly write the Lagrangians of point and gauge invariant interactions of higher-spin baryon resonances with nucleons. To this end, we study the general algebraic properties of coupling matrices preserving required symmetries. In Secs.~\ref{constraints on kernels} a basis set of such tensor matrices is determined. It is shown that there are only three basis matrices for any spin of the resonance. Then we construct minimally local Lagrangians of the interactions of baryon resonances with nucleons and pions and with nucleons and vector fields such as photons and vector mesons. We find that the symmetry classifies three terms of the Lagrangian in terms of the chiralities of the baryons involved. The classification and the definition of the nucleon-to-resonance form factors is in accord with perturbative QCD (PQCD) at asymptotic momentum transfers---the chirality (non)conserving form factors are asymptotically proportional to corresponding helicity (non)conserving amplitudes. This means that form factors exhibit simple power-law scaling, while form factors of lower symmetry can potentially mix different-scale helicity conserving and nonconserving contributions.

Section~\ref{fit section} concerns the point and gauge interactions of baryon resonances on the mass shell. The $Q^2$ dependencies of the Lagrangian form factors are parametrized in the vector dominance model developed in the papers~\cite{2007EPJA...34..223V, vereshkov:073007}. The point- and gauge-invariant form factors exhibit peculiar gross features in their $Q^2$ behavior. In particular, the data for $N\Delta(1232)$ allow for low-$Q^2$ scaling of the form factor ratios similar to the analogous phenomena observed in the case of elastic form factors \cite{PhysRevLett.91.092003}. In addition, the form factors for the transitions $pN(1520)$ and $pN(1680)$ are proportional to each other.


\section{\label{RS formulation}Constraints and symmetries in the RS theory}

Describing baryons of spin $J = \ell + \frac12 \geqslant \frac32$ as the RS fields $\Psi_{\mu_1\dots\mu_\ell}$ of the tensor rank $\ell$, we have to deal with the problems arising from the reducibility of the RS representations. The field $\Psi_{\mu_1\dots\mu_\ell}$ contains $N_\text{RS}$ components, which exceeds the number of spin-$J$ DOF $N_\text{DOF} = 2(2J+1) = 4(\ell+1)$:
\begin{align}
	N_\text{RS} = 4 {3+\ell \choose \ell} = N_\text{DOF} \left(1+\frac{\ell}2\right) \left(1+\frac{\ell}3\right).
\end{align}
To eliminate redundant DOF, we have to impose $\frac23 \ell (\ell+1) (\ell+5)$ constraints on the field, four-transversality and $\gamma$-tracelessness conditions \cite{ra-sh}:
\begin{gather}\label{eq:constraints}
	\partial^\lambda \Psi_{\lambda\mu_2\dots\mu_\ell}
	 = 0 = \gamma^\lambda \Psi_{\lambda\mu_2\dots\mu_\ell}.
\end{gather}
These constraints are linked to the symmetries of the RS theory that we illustrate by the simplest case of a vector-spinor field.


\subsection{\label{Covariant constraints}RS theory of a vector-spinor field}


The general Lagrangian for a vector-spinor field \cite{moldauer-case, 2004pilling, *2005pilling} can be written as
\begin{align}\label{free Lagrangian}
	\mathscr{L} &{}=
	  \bar\Psi^\mu \left[ i \Gamma_{\mu\nu\lambda}(A) \partial^\lambda
	                      - M \Gamma_{\mu\nu}(A) \right] \Psi^\nu
	{}- \bar\Psi_\mu J^\mu +\text{H.c.}
\end{align}
Here it is supposed that the interaction Lagrangian is linear in the RS field; i.e., the source $J_\mu$ depends only on external fields. The matrices in the kinetic and mass terms of the Lagrangian \eqref{free Lagrangian} are given by
\begin{align*}
	\Gamma_{\mu\nu\lambda}(A) ={}&
		g_{\mu\nu} \gamma_\lambda
		- A \gamma_\mu g_{\nu\lambda} -A^* \gamma_\nu g_{\mu\lambda}
		\\&
		+ \left( \frac32 \lvert A \rvert^2 - \Re A + \frac12 \right)
		\gamma_\mu \gamma_\lambda \gamma_\nu,
	\\
	\Gamma_{\mu\nu}(A) ={}&
		g_{\mu\nu} -\left( 3 \lvert A \rvert^2 - 3 \Re A + 1 \right)
		\gamma_\mu \gamma_\nu,
\end{align*}
where $A \neq \frac12$ is an arbitrary complex parameter.

Evaluating the Euler-Lagrange equation for the Lagrangian \eqref{free Lagrangian} gives the RS equation
\begin{align}\label{RS equation}
	\left[ i \Gamma_{\mu\nu\lambda}(A) \partial^\lambda - M \Gamma_{\mu\nu}(A) \right] \Psi^\nu = J_\mu.
\end{align}

Operating now on Eq.~\eqref{RS equation} by $\gamma^\mu$ and $\partial^\mu$, we obtain constraints
\begin{subequations}\label{constraints}
\begin{align}
	2 i \Phi + \left( 1 - 3A \right) i \nt\partial \Psi
	+ 3M \left( 1 - 2A \right) \Psi = \frac{1}{1-2 A^* } \gamma_\mu J^\mu,
\end{align}
\begin{multline}
	i \left( 1 - A^* \right) \nt\partial \Phi
	+ i \frac12 \left( 1 - A^* \right) \left( 1 - 3A \right) \partial^2 \Psi
	\\
	- M \Phi + M \left( 3 \lvert A \rvert^2 - 3 \Re A + 1 \right) \nt\partial \Psi
	= \partial_\mu J^\mu,
\end{multline}
\end{subequations}
where $\Phi = \partial^\mu \Psi_\mu$, and $\Psi = \gamma^\mu \Psi_\mu$. 

\subsubsection{Constraints and symmetries for a free vector-spinor field}

For the free field ($\partial^\mu J_\mu = 0 = \gamma^\mu J_\mu$) solving constraints \eqref{constraints} with respect to $\Psi$ and $\Phi$, we get the requisite conditions \eqref{eq:constraints} eliminating 8 redundant DOF:
\begin{align}\label{eq:3/2 constraints}
	\partial^\mu \Psi_\mu = 0 = \gamma^\mu \Psi_\mu.
\end{align}
These conditions recast Eq.~\eqref{RS equation} into the Dirac form
\begin{align}\label{reduced RS equation}
	\bigl( i \nt\partial - M \bigr) \Psi_\mu = 0.
\end{align}

We therefore are led to a well-known conclusion that the free RS theory is derived from a one-parameter equivalent class of Lagrangians. The free-field parameter $A$ is shifted to a value $A'$ by a field transformation 
\begin{align}\label{point transformation}
	\Psi'_\mu = \Theta_{\mu\nu} (A,A') \Psi^\nu,
\end{align}
where tensor matrix $\Theta_{\mu\nu}(A,A')$ is given by
\begin{align}
	\Theta_{\mu\nu} (A,A') =
		g_{\mu\nu} + \frac{A'-A}{2\left( 2 A - 1 \right)} \gamma_\mu \gamma_\nu.
\end{align}
The transformation \eqref{point transformation} is usually referred to as the ``point" or ``contact" transformation. Such transformations form a non-unitary symmetry group of the equivalent class of the free-field Lagrangians with the product law
\begin{gather}
	\Theta_{\mu}{}^{\eta} (A,A'') \Theta_{\eta\nu} (A'',A') = \Theta_{\mu\nu} (A,A').
\end{gather}
It should be stressed again that we use the definition of point transformations \eqref{point transformation} \textit{without} a simultaneous change of the free-field parameter $A' \to A$. Therefore, the equivalent class of the Lagrangians is point invariant, while a Lagrangian for any particular $A$ is not point invariant.

Besides, the free massless RS field admits the gauge transformation \cite{ra-sh,2004pilling,*2005pilling}
\begin{gather}\label{gauge transformations}
	\Psi'_\mu = \Psi_\mu + \partial_\mu \theta(x),
\end{gather}
where $i \nt\partial\theta = 0$.


\subsubsection{Constraints and symmetries for a vector-spinor field in the presence of interactions}


Introducing interactions to the theory complicates the issue of lower-spin states, because interaction Lagrangians could bring the components of the field into nontrivial mixing, therefore violating off-shell the equivalence of the theories with different values of the parameter $A$, breaking free-field constraints, and making nonphysical lower-spin states contribute to the observables \cite{ve-zw, johnson-sudarshan, 1973singh, 1998pascalutsa, 1989benmerrouche, 1971nath-etemadi, da-91, 1968peccei, 1969peccei, 1980nath-bhattacharyya}.

The most obvious way to prevent such inconsistencies is to impose free-field intrinsic symmetries on the interaction Lagrangian. Indeed, as we can see from Eqs.~\eqref{constraints}, the constraints in the general case of the interacting theory involve the operators $\partial^\mu J_\mu$ and $\gamma^\mu J_\mu$ depending on external fields. However, the requirement of the gauge and point invariance of the interactions makes these operators be identically zero:
\begin{align}\label{eq:Jconstr}
	\partial^\mu J_\mu = 0 = \gamma^\mu J_\mu.
\end{align}
Hereby for the point- and gauge-invariant interactions the free-field constraints \eqref{eq:3/2 constraints} remain unaffected and the field equation remains to be the Dirac equation for every vector component of the vector spinor field,
\begin{align}
	\bigl( i \nt\partial - M \bigr) \Psi_\mu = J_\mu.
\end{align}

It should be emphasized again that the requirements \eqref{eq:Jconstr} could be considered redundant from the mathematical point of view. It can be shown that requiring the interaction Lagrangian to be invariant under the gauge transformation \eqref{gauge transformations} solely still leads to a correct number of DOF present in the theory \cite{1998pascalutsa, pa-ti}. Indeed, for the gauge invariant interactions ($\partial^\mu J_\mu = 0$) and $A=1$, a field redefinition
\begin{align}
	\Psi_\mu = \Psi'_\mu + \frac{1}{3M} \partial_\mu \nt\partial \frac{1}{\partial^2} \gamma^\lambda J_\lambda
\end{align}
recasts the free-field Lagrangian \eqref{free Lagrangian} into the form
\begin{align}\label{eq:2.15}
	\mathscr{L} &{}=
	  \bar\Psi'{}^\mu ( i \nt\partial - M ) \Psi'_\mu
	{}- \bar\Psi'_\mu J^\mu +\text{H.c.},
\end{align}
while the constraints \eqref{constraints} become $\partial^\mu \Psi'_\mu = 0 = \gamma^\mu \Psi'_\mu$. The Lagrangian \eqref{eq:2.15} does not contain any new contact terms that could be associated with the interactions of the spurious lower-spin components of the field $\Psi_\mu$ \cite{2001PhLB..503...85P, 2009PhRvC..80b8201K}. Therefore, the gauge-invariant interactions do exclude spin-$\frac12$ DOF.

We choose, however, to amplify the symmetry of the interaction Lagrangian by the point invariance suggested by Peccei \cite{1968peccei}. The point invariance is an additional symmetry that constrains ambiguities in the definition of the gauge-invariant interaction Lagrangian and corresponding form factors. For example, consider the first possible differential order of the gauge-invariant $NRV$ Lagrangian for spin-3/2 resonance. It can contain at least two invariants $I_1 = \bar\Psi^{\mu\nu} N V_{\mu\nu}$ and $I_2 = i\bar\Psi^{\mu\nu} e_{\mu\nu\lambda\sigma} \gamma_5 N V^{\lambda\sigma}$. Arbitrary linear combinations of the invariants could be used as terms of the Lagrangian. For higher spins and higher derivatives, the number of invariants increases dramatically. At this point, we have two possibilities. The first one is to choose some couplings ``by hand.'' The second one is to expand the symmetry so that to constrain the invariants. The most obvious additional symmetry for RS fields is the point invariance, and it does leave only one invariant, as we will see. In addition to providing unambiguous definition of the Lagrangian and form factors, the point invariance leads to a classification of the form factors in line with the theory of the elastic form factors.


\subsection{Symmetries of higher-spin RS fields ($J \geqslant \frac52$)}

The theory of the free massive RS fields $\Psi_{\mu_1\dots\mu_\ell}$ for $J = \ell + \frac12 \geqslant \frac52$ was constructed by Singh and Hagen in Ref.~\cite{1974PhRvD...9..910S}. To get the necessary constraints \eqref{constraints}, it relies on introducing lower-spin auxiliary fields in the Lagrangian, which is mathematically inevitable for $\ell \geqslant 2$ \cite{1979berends}. The Singh-Hagen Lagrangians imply that the $\gamma$ tracelessness of the field ($\gamma^\lambda \Psi_{\lambda\mu_2\dots\mu_\ell} = 0$) does not follow from the field equation, but is an \textit{a priori} condition. In such a formulation the point invariance means that the Lagrangian is invariant under all nonsingular linear transformations of the auxiliary fields. Since the interaction source $J_{\mu_1\mu_2\dots\mu_\ell}$ is a variational derivative of the interaction part of the action, the $\gamma$ tracelessness of the source follows directly from the $\gamma$ tracelessness of the field:
\begin{align}\label{eq:Jtra}
	\gamma^\lambda J_{\lambda\mu_2\dots\mu_\ell} = 0,
	\quad
	J_{\mu_1\mu_2\dots\mu_\ell} = \frac{\delta}{\delta \bar\Psi^{\mu_1\mu_2\dots\mu_\ell}}\int \mathscr{L}_\text{int} \mathrm{d}^4 x.
\end{align}

It should be noted that at least for the spin-$\frac52$ we can construct free-field theory in lines with the spin-$\frac32$ theory, with $\gamma$ tracelessness of the field following from the field equation \cite{2010PhRvC..82a5203S, 2010arXiv1006.3929V}. In this case, to get the condition \eqref{eq:Jtra}, we should again require the invariance of the interaction Lagrangian under the point transformations of the RS field: 
\begin{gather}
	\Psi_{\mu_1\dots\mu_\ell}' = \Psi_{\mu_1\dots\mu_\ell}
	+ \frac{1}{\ell} \sum_{a=1}^\ell \gamma_{\mu_a} \theta_{\mu_1\dots\mu_{a-1}\mu_{a+1}\dots\mu_\ell},
	\\
	\gamma^{\mu_2} \theta_{\mu_2\dots\mu_\ell} = 0 = \partial^{\mu_2} \theta_{\mu_2\dots\mu_\ell}.
\end{gather}

Finally, in any case the constraint $\partial^\lambda \Psi_{\lambda\mu_2\dots\mu_\ell} = 0$ is guaranteed by the gauge invariance of the interaction Lagrangian and the kinematic term of the free Lagrangian:
\begin{gather}
	\Psi_{\mu_1\dots\mu_\ell}' = \Psi_{\mu_1\dots\mu_\ell}
	+ \frac{1}{\ell} \sum_{a=1}^\ell \partial_{\mu_a} \theta_{\mu_1\dots\mu_{a-1}\mu_{a+1}\dots\mu_\ell},
	\\
	\nt\partial \theta_{\mu_2\dots\mu_\ell} = 0,
	\qquad
	\gamma^{\mu_2} \theta_{\mu_2\dots\mu_\ell} = 0 = \partial^{\mu_2} \theta_{\mu_2\dots\mu_\ell}.
\end{gather}


\section{\label{lagrangians}Point- and gauge-invariant nucleon-resonance interactions}

The concern of this section is explicit algebraic construction of the point- and gauge-invariant $NR$ interactions ($N$ denotes a nucleon and $R$ a spin-$J$ resonance) with pions and vector fields, e.g., photons, $\rho$ and $\omega$ mesons.


\subsection{\label{constraints on kernels}Symmetry constraints on coupling matrices}

The invariance of the interactions under the gauge transformations \eqref{gauge transformations} of the RS field implies that the interaction Lagrangian should be a functional of the gauge-invariant curvature $\Psi_{([\mu_1\nu_1][\mu_2\nu_2]\dots[\mu_\ell\nu_\ell])}$:
\begin{align}
	\mathscr{L}_\text{int}
	= {}&\mathscr{L}_\text{vect} \bigl\{ \Psi_{\bar{A}}(x), V_{\mu\nu}(x'), N(x'') \bigr\}
	\notag\\&{}+ \mathscr{L}_\text{pion} \bigl\{ \Psi_{\bar{A}}(x), \pi(x'), N(x'') \bigr\},
\end{align}
where $\bar{A} = ([\mu_1\nu_1][\mu_2\nu_2]\dots[\mu_\ell\nu_\ell])$ is a multi-index and $V_{\mu\nu}$ is a strength tensor of a vector field $V_\mu$. All fields are assumed to be isotopic scalars for brevity. The square brackets $[\cdot\cdot]$ enclose antisymmetric pairs of indices and the parentheses $(\cdots)$ indicate that the tensor spinor is symmetric under permutations of the pairs $[\mu_a\nu_a]$.

For $\ell=1$ and 2 the explicit form of the curvature $\Psi_{\bar{A}}$ coincides with the Maxwell field strength and the linearized Riemann curvature tensor, while for higher spins the curvature is easily constructed as a suitably (anti)symmetrized $\ell$-th derivative of the RS field\footnote{There is another definition of the generalized curvatures $\Psi_{(\mu_1\dots\mu_\ell)(\nu_1\dots\nu_\ell)}$ in the literature \cite{PhysRevD.21.358}. However, the curvature $\Psi_{(\mu_1\dots\mu_\ell)(\nu_1\dots\nu_\ell)}$ of Ref.~\cite{PhysRevD.21.358} can be represented as a linear combination of the curvature $\Psi_{([\mu_1\nu_1][\mu_2\nu_2]\dots[\mu_\ell\nu_\ell])}$ that is used in this paper.}:
\begin{align} \label{eq:3.1}
	&\ell = 1: \quad \Psi_{[\mu_1\nu_1]} = \Psi_{\nu_1,\mu_1} - \Psi_{\mu_1,\nu_1};
	\\ \notag
	&\ell = 2: \quad \Psi_{([\mu_1\nu_1][\mu_2\nu_2])} = \frac12 \bigl( \Psi_{\nu_1\nu_2,\mu_1\mu_2} 
	\\&\phantom{\ell = 2: \quad }
	{} - \Psi_{\nu_1\mu_2,\mu_1\nu_2} - \Psi_{\mu_1\nu_2,\nu_1\mu_2} + \Psi_{\mu_1\mu_2,\nu_1\nu_2} \bigr);
	\\ \notag  \label{eq:3.3}
	&\ell \geqslant 3: \quad \Psi_{([\mu_1\nu_1][\mu_2\nu_2]\dots[\mu_\ell\nu_\ell])}
		\\&\phantom{\ell = 2: \quad }
	{} = \frac1{2^{\ell-1}} \left( \Psi_{\nu_1\nu_2\dots\nu_\ell,\mu_1\mu_2\dots\mu_\ell} + \dots \right).
\end{align}

The effective interactions of composite particles are basically nonlocal and the Lagrangian corresponding to such interactions can contain all the couplings (despite the order of derivatives involved) preserving required symmetries
\begin{align}
	\mathscr{L}_\text{vect} ={}&
	\sum_{i,k=0}^{+\infty} \frac{i^{\ell+i+k} g_{i,k}}{2M_N^{\ell+i+k+1}} \bar\Psi^{\bar{A},\bar\alpha} \Gamma_{\bar{A}\bar\alpha\bar\beta\lambda\sigma} \gamma_R N^{,\bar\beta} V^{\lambda\sigma}
	+ \text{H.c.},
	\label{LV}
	\\
	\mathscr{L}_\text{pion} ={}&
	\sum_{i,k=0}^{+\infty} \frac{i^{\ell+i+k}f_{i,k}}{2M_R^{\ell+i} m_\pi^k}
	\bar\Psi^{\bar{A},\bar\alpha} \Gamma_{\bar{A}\bar\alpha\bar\beta} \gamma_R \gamma_5 N \pi^{,\bar\beta}
	+ \text{H.c.},
	\label{Lpi}
\end{align}
where $\bar\alpha=\alpha_1\alpha_2 \dots \alpha_i$ and $\bar\beta=\beta_1\beta_2 \dots \beta_k$ are multi-indices coming from the derivatives of field operators, $\gamma_R = 1$ for baryon resonances with spin-parities $J^P = \frac32^-$, $\frac52^+\dots$, and $\gamma_R = i\gamma_5$ in other cases.

Each term of the above derivative expansion of the interaction Lagrangian comprises coupling matrices $\Gamma_{\bar{A}\bar\alpha\bar\beta}$. The point invariance of the interaction Lagrangian, which is equivalent to the $\gamma$ tracelessness \eqref{eq:Jtra} of the current $J_{\mu_1\dots\mu_\ell}$, results in the couplings $\Gamma_{\bar{A}\bar\alpha\bar\beta}$ being $\gamma$-traceless; i.e.,
\begin{align}\label{gamma-transversality}
	\gamma^{\mu_1} \Gamma_{([\mu_1\nu_1][\mu_2\nu_2]\dots[\mu_\ell\nu_\ell])\bar\alpha\bar\beta}= 0.
\end{align}

Thereby, to implement the idea of gauge and point invariance of the interactions, one should find all the (pseudo)tensor matrices satisfying Eq.~\eqref{gamma-transversality}. To this end, we write out the general expression for the matrix $\Gamma_{\bar A\bar\alpha}$ as a decomposition in terms of a complete orthonormal basis set made up of the identity matrix and matrices $\gamma_5$, $\gamma_\mu$, $i\gamma_\mu \gamma_5$, and $i\sigma_{\mu\nu}$\footnote{In this paper we use a convention $\gamma_5 = \frac{i}{4!} e_{\mu\nu\lambda\sigma} \gamma^\mu \gamma^\nu \gamma^\lambda \gamma^\sigma$, $\sigma_{\mu\nu} = \frac12 \left( \gamma_\mu \gamma_\nu - \gamma_\nu \gamma_\mu \right)$, $g_{00} = +1$ and $g_{ii} = -1$, $e_{0123} = -1$, $\bar\Gamma = \gamma_0 \Gamma^\dagger \gamma_0$.}
\begin{align}
	\Gamma_{\bar{A}\bar\alpha}
	=
	\frac14 \Tr \bigl[ \Gamma_{\bar{A}\bar\alpha} \bigr]
	+ \frac14 \gamma_5 \Tr \bigl[ \gamma_5 \Gamma_{\bar{A}\bar\alpha} \bigr]
	+ \dots,
\end{align}
with the traces being considered as (pseudo)tensor coefficients of the decomposition. Then imposing the requirement~\eqref{gamma-transversality} on the decomposition constrains the coefficients. These calculations are carried out in the Appendix~\ref{app:a}. Here we proceed to the results.

The most general $\gamma$-traceless matrix of any tensor rank can be decomposed in terms of explicitly traceless basis as follows:
\begin{align}\label{total decomposition}
	\Gamma_{\bar A\bar\alpha}
	 = {}&\Gamma_{\bar A \bar B} \left[ \mathsf{R}^{\bar B}{}_{\bar\alpha}
	   + \gamma_\rho \mathsf{S}^{\rho\bar B}{}_{\bar\alpha}
	+  \sigma_{\rho\omega} \mathsf{T}^{[\rho\omega]\bar B}{}_{\bar\alpha}
	\right],
\end{align}
where $\mathsf{R}_{\bar B\bar\alpha}$, $\mathsf{S}_{\bar B\bar\alpha}$, and $\mathsf{T}_{\bar B\bar\alpha}$ are some coefficient tensors. In the simplest case of $\ell=1$ the matrix $\Gamma_{\bar A \bar B} = \Gamma_{[\mu\nu][\lambda\sigma]}$ can be written as
\begin{align}\label{eq:kernel1}
	\Gamma_{[\mu\nu][\lambda\sigma]} 
	= -\frac16 \left( \sigma_{\mu\nu} \sigma_{\lambda\sigma}
	                 +3 \sigma_{\lambda\sigma} \sigma_{\mu\nu} \right).
\end{align}
For higher $\ell$ the matrix $\Gamma_{\bar A \bar B}$ is defined by the following recurrence relation:
\begin{align}
\Gamma_{\bar A \bar B}^{(\ell)}
	&{} = \frac{3}{2(2\ell+1)\ell^2}\sum_{a,b=1}^\ell \biggl[ (\ell+1) \Gamma^{(\ell-1)}_{\bar A^a \bar B^b} \Gamma^{(1)}_{A^a B^b}
	+{}
		\notag \\ \label{eq:kernel_ell}
	&{}+(\ell-1) \Gamma^{(\ell-1)}_{\bar A^a \bar B^b} \Gamma^{(1)}_{B^b A^a}
	- \sum_{b \ne c=1}^{\ell} \Gamma^{(\ell-1)}_{\bar A^a A^a\bar B^{bc}}
	\Gamma^{(1)}_{B^b B^c} \biggr],
\end{align}
where we have introduced a shorthand notation for the multi-indices:
\begin{equation}\label{eq:indices}
\begin{gathered}
	A^a = [\mu_a\nu_a], \qquad \bar A = (A^1 \dots A^\ell),
	\\
	\bar A^a = (A^1 \dots A^{a-1} A^{a+1} \dots A^\ell),
	\\
	B^b = [\lambda_b\sigma_b], \qquad \bar B = (B^1 \dots B^\ell),
	\\
	\bar B^b = (B^1 \dots B^{b-1} B^{b+1} \dots B^\ell).
\end{gathered}
\end{equation}

The matrices \eqref{eq:kernel1} and \eqref{eq:kernel_ell} are traceless and self-conjugate:
\begin{align}
	\gamma^{\mu_1} \Gamma_{\bar A \bar B} = 0 = \Gamma_{\bar A \bar B} \gamma^{\lambda_1},
	\qquad
	\bar\Gamma_{\bar A \bar B} = \Gamma_{\bar B \bar A}.
\end{align}
The normalization of the matrix $\Gamma_{\bar A \bar B}$ is chosen so that
\begin{align}\label{eq:norm}
	\frac1{(q^2)^\ell} \Gamma_{\bar A \bar B} \prod_{i=1}^\ell q^{\nu_i} q^{\sigma_i}
	= P^{(\ell+\frac12)}_{\bar\mu\bar\lambda} (q),
\end{align}
where $\bar{\mu}=\mu_1\mu_2...\mu_\ell$, $\bar{\lambda}=\lambda_1\lambda_2...\lambda_\ell$; $P^{(\ell+\frac12)}_{\bar\mu\bar\lambda} (q)$ is a projector on the pure spin-$\bigl(\ell+\frac12 \bigr)$ state defined in Refs.~\cite{1957PhRv..106..345B, springerlink:10.1007/BF02747684}.

To clarify Eqs.~\eqref{total decomposition}--\eqref{eq:kernel_ell}, we should comment on several points. First of all, the importance of Eqs.~\eqref{total decomposition} needs to be stressed. These equations significantly simplify constructing all possible coupling matrices of the point- and gauge-invariant interactions, since they reduce the problem to finding all the tensor coefficients made up of the structure tensors of the Lorentz group $g_{\mu\nu}$ and $e_{\mu\nu\lambda\sigma}$. Constructing such tensor objects is a much easier problem mathematically compared with constructing tensor matrices.

An additional point to emphasize is the simplicity of the results obtained---there are only three traceless tensor matrices in the basis:
\begin{align}\label{basis set}
	\Gamma_{\bar{A}\bar{B}},
	\qquad
	\Gamma_{\bar{A}\bar{B}} \gamma_\rho,
	\qquad
	\Gamma_{\bar{A}\bar{B}} \sigma_{\rho\omega}.
\end{align}
It should be noted that the $\gamma$-traceless matrices \eqref{basis set} do not form a basis in the strict mathematical sense, for not all of their components are independent; i.e., the following identity holds:
\begin{align}\label{i1}
	 \Gamma_{\bar{A}\bar{B}^\ell[\lambda_\ell\sigma_\ell]} \gamma_\rho
	+\Gamma_{\bar{A}\bar{B}^\ell[\rho\lambda_\ell]} \gamma_{\sigma_\ell}
	+\Gamma_{\bar{A}\bar{B}^\ell[\sigma_\ell\rho]} \gamma_{\lambda_\ell} = 0.
\end{align}

Besides, a peculiar feature of the model (following from the symmetries of the Lagrangian) is that for a particular spin $J_R$ all the basis elements \eqref{basis set} are expressed solely by means of the matrix $\Gamma_{\bar{A}\bar{B}}$, which, in turn, is expressed through the simplest $\gamma$-traceless matrix $\Gamma_{[\mu\nu][\lambda\sigma]}$ defined by Eq.~\eqref{eq:kernel1}. The tensor matrix \eqref{eq:kernel1} therefore occurs in all point- and gauge-invariant interaction Lagrangians. In constructing such interactions $\Gamma_{[\mu\nu][\lambda\sigma]}$ assumes the similar role as the matrix $\gamma_\mu$ in the general case.


\subsection{\label{explicit Lagrangian}Minimal local interaction Lagrangian}

The vertex for the $NR$ interaction should involve three invariant amplitudes for the interactions of higher-spin resonances ($J_R \geqslant 3/2$) with nucleons and vector fields and one amplitude for the interactions with nucleons and pions. These amplitudes are nonlocal in the general case---they are given by Taylor expansions in the momentum space or derivative expansions \eqref{LV} and \eqref{Lpi} in the configuration space. However, practical calculations of electromagnetic $NR$ form factors are usually performed by utilizing minimally localized Lagrangians that contain a minimal possible number of derivatives and lead to minimal possible degrees of kinematic invariants in the observables. Based on the knowledge of the lowest-rank $\gamma$-traceless matrices \eqref{basis set}, we can now write minimally local interaction Lagrangians that are invariant under the point and gauge transformations of the RS field.

The minimally local Lagrangian of $NR\pi$ interactions could be written most easily \cite{2009PhRvC..80e8201S, 2010PhRvC..82a5203S}. It should involve $3 \ell$ derivatives of the field operators, since we have learned earlier that the simplest traceless coupling matrix for the spin $J_R=\ell+\frac12$ is the tensor matrix $\Gamma_{\bar A\bar B}$ \eqref{eq:kernel1} of the tensor rank $4 \ell$. Therefore, we have
\begin{align}\label{eq:NRpi}
	\mathscr{L}_{NR\pi}^{(\ell)}
	=
	\frac{i^{\ell}f}{2M_R^{2\ell} m_\pi^\ell}
	\bar\Psi^{\bar A,\lambda_1\dots\lambda_\ell} \Gamma_{\bar A\bar B} \gamma_R \gamma_5
	N \pi^{,\sigma_1\dots\sigma_\ell},
\end{align}
where $\gamma_R = 1$ and $i\gamma_5$ for the resonances $R$ with the parity $P = (-1)^\ell$ and $P = (-1)^{\ell+1}$, respectively; the multi-indices $\bar A = ([\mu_1\nu_1] \dots [\mu_\ell\nu_\ell])$ and $\bar B = ([\lambda_1\sigma_1] \dots [\lambda_\ell\sigma_\ell])$ follow the notation of Eq.~\eqref{eq:indices}. The $NR\pi$ Lagrangian \eqref{eq:NRpi} was first written in terms of the projectors $P^{(\ell+\frac12)}_{\bar\mu\bar\lambda} (\partial)$ given by Eq.~\eqref{eq:norm} in Refs.~\cite{2009PhRvC..80e8201S, 2010PhRvC..82a5203S}. 

It is worth pointing out that the minimally local point- and gauge-invariant Lagrangian \eqref{eq:NRpi} is defined unambiguously by the symmetry; i.e., there is no other Lagrangian term with the minimal number of the derivatives of the field operator.

In the same way as in the case of the $NR\pi$ Lagrangian, the first term in the derivative expansion of the point- and gauge-invariant $NRV$ Lagrangian \eqref{LV} for $J_R = \ell + 1/2$ is determined exactly:
\begin{align}\label{eq:L1}
	\mathscr{L}_{1}^{(V,\ell)}
	=
	\frac{i^{\ell}g_1^V}{2M_N^{3\ell-1}}
	\bar\Psi^{\bar A,\lambda_2\dots\lambda_\ell}
	\Gamma_{\bar A\bar B} \gamma_R N^{,\sigma_2\dots\sigma_\ell}
	V^{\lambda_1\sigma_1},
\end{align}
where $V = \gamma$, $\rho(770)$, $\omega(782)$,... is a photon or vector-meson field.

The choice of the next two $N R V$-vertices in Eq.~\eqref{LV} is more subtle. There are two coupling matrices of the $(4\ell+1)$st tensor rank:
\begin{align}
	\Gamma_{\bar A\bar B} \gamma_\rho \gamma_R,
	\qquad
	\Gamma_{\bar{A}\bar{B}^\ell[\rho\lambda_\ell]} \gamma_{\sigma_\ell} \gamma_R
	-\Gamma_{\bar{A}\bar{B}^\ell[\rho\sigma_\ell]} \gamma_{\lambda_\ell} \gamma_R.
\end{align}
However, only one of the above tensor matrices is linearly independent owing to the identity \eqref{i1}. Thus, we can write the second term of the Lagrangian as
\begin{align}\label{3rd-order L}
	\mathscr{L}_{2}^{(V,\ell)}
	={}&
	\frac{i^{\ell+1}g_2^V}{2M_N^{3\ell-1} M_R}
	\bar\Psi^{\bar A,\lambda_2\dots\lambda_\ell\rho}
	\Gamma_{\bar A\bar B} \gamma_\rho \gamma_R N^{,\sigma_2\dots\sigma_\ell}
	V^{\lambda_1\sigma_1}.
\end{align}
The general point- and gauge-invariant $NRV$ Lagrangian with $3\ell+1$ derivatives could contain another term that differs from the Lagrangian \eqref{3rd-order L} by a place of the $(3\ell+1)$st derivative $\partial_\rho$:
\begin{align}
	\mathscr{L}_{(2)}'
	=
	\frac{i^{\ell+1}g'_2{}^V}{2M_N^{3\ell-1} M_R}
	\bar\Psi^{\bar A,\lambda_2\dots\lambda_\ell}
	\Gamma_{\bar A\bar B} \gamma_\rho \gamma_R N^{,\sigma_2\dots\sigma_\ell\rho}
	V^{\lambda_1\sigma_1}.
\end{align}
This term can be converted to the $(3\ell)$th order $NRV$ Lagrangian \eqref{eq:L1} by virtue of the on-shell field equations for the nucleon field. If the nucleon is off its mass shell, then the Lagrangian $\mathscr{L}_{(2)}'$ will lead to the third degree of the invariant mass of the nucleon $M_N$ in the observables, while $\mathscr{L}_{(2)}$ will give the first degree. Thus, the vertex $\mathscr{L}_{(2)}'$ should be omitted from the minimally local Lagrangian, as it represents a next-order term of the nonlocal derivative expansion \eqref{LV}, whose leading term is the Lagrangian \eqref{eq:L1}.

Finally, to find the third term of the Lagrangian, we have to consider the terms with another additional derivative. Using the results of Sec.~\ref{constraints on kernels}, it is easy to show that there are four linearly independent matrices of the $(4\ell+2)$nd tensor rank:
\begin{gather}\label{eq:kernels6}
	\Gamma_{\bar A\bar B} \gamma_R g_{\rho\omega},
	\qquad
	\Gamma_{\bar A\bar B} \sigma_C \gamma_R,
	\qquad 
	\Gamma_{\bar A\bar B^1[\rho\omega]} \sigma_{B^1} \gamma_R,
	\\ 
	\bigl( \Gamma_{\bar A\bar B^1[\lambda_1\rho]} g_{\sigma_1\omega}
	-\Gamma_{\bar A\bar B^1[\sigma_1\rho]} g_{\lambda_1\omega}
	+\Gamma_{\bar A\bar B^1[\lambda_1\omega]} g_{\sigma_1\rho}
	\notag\\ \label{eq:sym_m}
	-\Gamma_{\bar A\bar B^1[\sigma_1\omega]} g_{\lambda_1\rho}
	-\Gamma_{\bar A\bar B^1[\lambda_1\sigma]} g_{\rho\omega} \bigr) \gamma_R.
\end{gather}
The matrix \eqref{eq:sym_m} is chosen so that it is symmetric and traceless in $\rho\omega$.

In the general case, we can write five Lagrangian terms containing the matrices \eqref{eq:kernels6} and \eqref{eq:sym_m}. Obviously, only one of these terms gives the third vertex and other four are next-order corrections to the first two Lagrangian terms \eqref{eq:L1} and \eqref{3rd-order L}. It can be shown that it is the matrix \eqref{eq:sym_m} that gives the third independent term:
\begin{align}
	\mathscr{L}_3^{(V,\ell)} = {}&
	\frac{i^{\ell}g_3}{2M_N^{3\ell-1}M_R^2}
	\bar\Psi^{\bar A,\lambda_2\dots\lambda_\ell\rho}
	\bigl(
		 \Gamma_{\bar A\bar B^1[\lambda_1\rho]} g_{\sigma_1\omega}
		\notag\\&{}
		-\Gamma_{\bar A\bar B^1[\sigma_1\rho]} g_{\lambda_1\omega}
		+\Gamma_{\bar A\bar B^1[\lambda_1\omega]} g_{\sigma_1\rho}
		\notag\\&{}
		-\Gamma_{\bar A\bar B^1[\sigma_1\omega]} g_{\lambda_1\rho}
		-\Gamma_{\bar A\bar B^1[\lambda_1\sigma_1]} g_{\rho\omega}
	\bigr) \gamma_R
		\notag\\&{}\times
	N^{,\sigma_2\dots\sigma_\ell\omega} V^{\lambda_1\sigma_1}.
\end{align}

Finally, the minimally local point- and gauge-invariant Lagrangian of the electromagnetic nucleon-resonance interactions can be written for the resonances of the spin $J_R = \ell + 1/2$ as follows:
\begin{gather}\label{eq:LRNV}
\mathscr{L}^{(\ell)} = \sum_V \left[ \mathscr{L}_1^{(V,\ell)} + \mathscr{L}_2^{(V,\ell)} + \mathscr{L}_3^{(V,\ell)} \right] + \text{ H.c.},
\\\label{lagrangian}
\begin{aligned}
	\mathscr{L}_1^{(V,\ell)} = {}&\frac{i^{\ell}g_1^V}{2M_N^{3\ell-1}} \bar\Psi^{\bar A,\lambda_2\dots\lambda_\ell} \Gamma_{\bar A\bar B} \gamma_R N^{,\sigma_2\dots\sigma_\ell} V^{\lambda_1\sigma_1},
 \\
	\mathscr{L}_2^{(V,\ell)} = {}&\frac{i^{\ell+1}g_2^V}{2M_N^{3\ell-1}M_R} \bar\Psi^{\bar A,\lambda_2\dots\lambda_\ell\rho} \Gamma_{\bar A\bar B} \gamma_\rho \gamma_R N^{,\sigma_2\dots\sigma_\ell} V^{\lambda_1\sigma_1},
 \\
	\mathscr{L}_3^{(V,\ell)} = {}&
	\frac{i^{\ell}g_3^V}{2M_N^{3\ell-1}M_R^2}
	\bar\Psi^{\bar A,\lambda_2\dots\lambda_\ell\rho}
	\bigl(
		 \Gamma_{\bar A\bar B^1[\lambda_1\rho]} g_{\sigma_1\omega}
		\notag\\&{}
		-\Gamma_{\bar A\bar B^1[\sigma_1\rho]} g_{\lambda_1\omega}
		+\Gamma_{\bar A\bar B^1[\lambda_1\omega]} g_{\sigma_1\rho}
		\notag\\&{}
		-\Gamma_{\bar A\bar B^1[\sigma_1\omega]} g_{\lambda_1\rho}
		-\Gamma_{\bar A\bar B^1[\lambda_1\sigma_1]} g_{\rho\omega_1}
	\bigr) \gamma_R
		\notag\\&{}\times
	N^{,\sigma_2\dots\sigma_\ell\omega} V^{\lambda_1\sigma_1},
\end{aligned}
\end{gather}
where the multi-indices $\bar A$, $\bar B$ are the same as in Eq.~\eqref{eq:indices}; $\gamma_R = 1$ for baryon resonances with spin-parities $J^P = \frac32^-$, $\frac52^+\dots$ and $\gamma_R = i\gamma_5$ in other cases.

We can see that the requirement of the point and gauge invariance leads to elegant and unified Lagrangians \eqref{eq:NRpi} and \eqref{eq:LRNV}---the interactions have the same coupling structure for arbitrarily high spin of the resonance. Each term of the Lagrangian comprises $(4\ell)$th tensor rank coupling matrix $\Gamma_{\bar A\bar B}$ defined by the recurrence relation \eqref{eq:kernel_ell}. As follows from Eq.~\eqref{eq:kernel_ell}, explicit form of $\Gamma_{\bar A\bar B}$ is a sum of $(\ell !)^3 (\ell+1)$ products of $2^\ell$ $\sigma$-matrices, which is a quite formidable expression for higher spins. However, this does not pose a problem to calculating observable quantities. Since lower-spin components of the RS field do not participate in the interactions, the propagator of higher-spin resonance in $S$-matrix elements of the point- and gauge-invariant theory is proportional to the projector $P^{\bar\alpha\bar\nu}_{(\ell+1/2)}(p)$. Thus, to calculate observables, we have to deal with products of the projectors and coupling matrices $P^{\bar\alpha\bar\nu}_{(\ell+1/2)}(p) \Gamma_{\bar A\bar B}$. In the Appendix~\ref{app:a} it is proven that  the tensor matrix $\Gamma_{\bar A\bar B}$ in such products for arbitrarily high $\ell$ can always be reduced to the simplest $\gamma$-traceless matrix $\Gamma_{[\mu\nu][\lambda\sigma]}$. In particular, for the $NRV$ vertex we have
\begin{multline}\label{eq:1}
	P^{\bar\alpha\bar\nu}_{(\ell+\frac12)}(p)
	\mathscr{K}^{(V,\ell)}_{\bar A\bar B}(p,k) \prod_{a=1}^\ell p^{\mu_a} q^{\sigma_a}
	\cdot \prod_{b=2}^\ell p^{\lambda_b}
	\\=
	(p^2)^{\ell-1} P^{\bar\alpha\bar\nu}_{(\ell+\frac12)}(p)
	\mathscr{K}^{(V,1)}_{[\mu\nu_1][\lambda_1\sigma]}(p,k) p^\mu q^\sigma \prod_{a=2}^\ell q^{\nu_a},
\end{multline}
where $\bar A = ([\mu_1\nu_1] \dots [\mu_\ell\nu_\ell])$, $\bar B = ([\lambda_1\sigma_1] \dots [\lambda_\ell\sigma_\ell])$ and $\bar\alpha = (\alpha_1 \dots \alpha_\ell)$, $\bar\nu = (\nu_1 \dots \nu_\ell)$ are multi-indices; the vertex matrices $\mathscr{K}^{(V,\ell)}_{\bar A\bar B}$ are defined as follows:
\begin{multline}\label{eq:2}
	\mathscr{K}^{(V,\ell)}_{\bar A\bar B}(p,k)
	= \Biggl[ g_1^V \Gamma_{\bar A\bar B}
	-\frac{g_2^V}{M_R} \Gamma_{\bar A\bar B} \nt{p}
	\\
	+\frac{g_3^V}{M_R^2} \bigl(
		\Gamma_{\bar A\bar B^1[\lambda_1\rho]} g_{\sigma_1\omega}
		-\Gamma_{\bar A\bar B^1[\sigma_1\rho]} g_{\lambda_1\omega}
		+\Gamma_{\bar A\bar B^1[\lambda_1\omega]} g_{\sigma_1\rho}
\\
		-\Gamma_{\bar A\bar B^1[\sigma_1\omega]} g_{\lambda_1\rho}
		-\Gamma_{\bar A\bar B^1[\lambda_1\sigma_1]} g_{\rho\omega}
	\bigr) p^\rho k^\omega \Biggr] \gamma_R.
\end{multline}

In the case of $NR\pi$ vertex we get an even simpler result---the coupling matrix $\Gamma_{\bar A\bar B}$ reduces to the projector owing to Eq.~\eqref{eq:norm} and abandons the vertex:
\begin{multline}\label{eq:3}
	P^{\bar\alpha\bar\nu}_{(\ell+\frac12)}(p)
	\Gamma_{\bar A\bar B} \prod_{a=1}^\ell p^{\mu_a} p^{\lambda_a} q^{\sigma_a}
	\\
	=
	(p^2)^{\ell} P^{\bar\alpha\bar\nu}_{(\ell+\frac12)}(p)
	P_{\bar\nu\bar\sigma}^{(\ell+\frac12)}(p)
	\prod_{a=1}^\ell q^{\sigma_a}
	\\
	= (p^2)^{\ell} P^{\bar\alpha\bar\nu}_{(\ell+\frac12)}(p) \prod_{a=1}^\ell q_{\nu_a}.
\end{multline}

\begin{figure}%
	\center\includegraphics[width=0.95\linewidth]{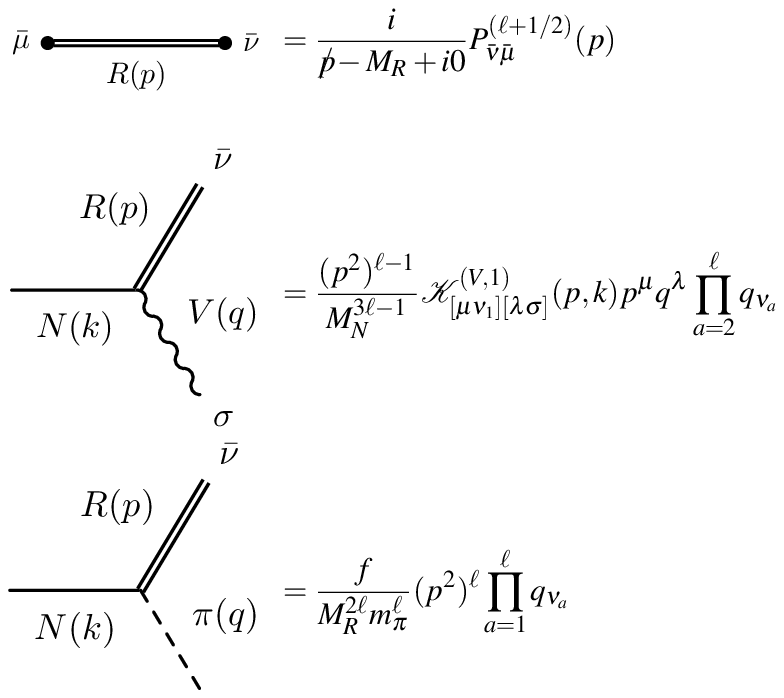}
	\caption{\label{fig:diagrams}Feynman rules for the propagator of a baryon resonance $R$ with spin $J_R=\ell+1/2$ and the $NRV$ and $NR\pi$ vertices ($\bar\mu=\mu_1\mu_2\dots\mu_\ell$, $\bar\nu=\nu_1\nu_2\dots\nu_\ell$). The vertex matrix $\mathscr{K}^{(V,1)}_{[\mu\nu][\lambda\sigma]}(p,k)$ is defined in Eq.~\eqref{eq:2}.}
\end{figure}

The relevant diagrams and Feynman rules making use of Eqs.~\eqref{eq:1} and \eqref{eq:3} are depicted in Fig.~\ref{fig:diagrams}.

\subsection{Helicity amplitudes for the transition of a nucleon to higher-spin resonance}

Using the Feynman rules depicted in Fig.~\ref{fig:diagrams}, we can calculate the cross section of resonant electroproduction in the gauge- and point-invariant model, the helicity amplitudes being of the form
\begin{multline}\label{A32}
	A^{(p,n)}_{3/2}(Q^2)
	= \mp \sqrt{N_\ell} \Bigl[ \left( Q^2 \pm \mu_\pm M_N \right) F_1^{(p,n)}(Q^2)
\\
	+ \mu_\pm M_R F_2^{(p,n)}(Q^2)
	- \left( Q^2+ \mu_\pm M_R \right) F_3^{(p,n)}(Q^2) \Bigr],
\end{multline}
\begin{multline}\label{A12}
	A^{(p,n)}_{1/2}(Q^2)
	= -\sqrt{\frac{\ell N_\ell}{\ell+2}} \Bigl[ \mu_\pm M_R F_1^{(p,n)}(Q^2)
\\
	+ \left( Q^2 \pm \mu_\pm M_N \right) F_2^{(p,n)}(Q^2)
		\mp \mu_\pm M_N F_3^{(p,n)}(Q^2) \Bigr],
\end{multline}
\begin{align}\label{S12}
	S^{(p,n)}_{1/2}&(Q^2)
	= \mp \sqrt{\frac{\ell N_\ell}{2(\ell+2)}} Q_+ Q_- \Biggl[ F_1^{(p,n)}(Q^2)
	\notag\\
	&{} - F_2^{(p,n)}(Q^2)
+ \frac{Q^2+M_R^2+M_N^2}{2M_R^2} F_3^{(p,n)}(Q^2) \Biggr],
\end{align}
where $N_\ell(Q^2) = \frac{\pi\alpha (\ell+2) Q_\pm^{2(\ell-1)} Q_\mp^{2\ell} M_R^{2(\ell-1)}}{2^{2\ell-1} \ell \tau_{\ell+1} M_N^{6\ell-1}(M_R^2-M_N^2)}$, $\tau_{\ell+1} = (2\ell+1)!!/(\ell+1)!$, $\mu_\pm = M_R \pm M_N$, and $Q_\pm = \sqrt{Q^2+\mu_\pm^2}$. In Eqs.~\eqref{A32}--\eqref{S12} the top and bottom signs correspond to the resonances with spin parities $J^P=(3/2)^{\pm},\ (5/2)^{\mp},\ ...$, respectively.

Note that in the helicity amplitudes \eqref{A32}--\eqref{S12} only normalization factors depend upon the spin of the resonance, while the polynomials accompanying form factors (FFs) are universal for arbitrarily high spin. This property of the model is a direct consequence of its symmetry---the point and gauge invariance of the interactions.

In the point- and gauge-invariant theory the FFs $F_f^{(p,n)}(Q^2)$ are given by
\begin{align}\label{FF series}
	F_f^{(p,n)}(Q^2) = \sum_{i=0}^{+\infty} F_{f,i}^{(p,n)} (Q^2) \left( \frac{Q^2}{4 M_N^2} \right)^i,
	\quad f=1,\,2,\,3.
\end{align}
Here the Taylor series in $Q^2$ comes from the higher terms of the derivative expansion \eqref{LV} of the interaction Lagrangian; this takes into account the effects of nonlocality of the nucleon and its resonances. The coefficients $F_{f,i}^{(p,n)}$ at powers of $Q^2$ are dispersionlike expansions specified by vector-meson--dominance model.

In what follows, however, we consider only the first term $F_{f,0}^{(p,n)}$ in the series \eqref{FF series}. The FFs $F_{f,0}^{(p,n)}$ correspond to the minimally local Lagrangian \eqref{eq:LRNV} and are sufficient to describe the observed on-shell $Q^2$ dependencies of the helicity amplitudes.


\subsection{On-shell equivalence of models of transition form factors}

For sufficiently high momentum transfers, there arises the question of differences between the point- and gauge-invariant model \eqref{eq:NRpi} and \eqref{eq:LRNV} and the alternatives of lower symmetry. The question is of importance, because different choices of the tensor-matrix kernels of the interaction Lagrangians may lead to differences in polynomials preceding FFs $F_f^{(N)}(Q^2)$ (compare the point- and gauge-invariant helicity amplitudes \eqref{A32}--\eqref{S12} with their counterparts from Refs.~\cite{Devenish1976, la-06, vereshkov:073007, pa-07}).

From general field-theoretical considerations, it is expected that both conventional and gauge-invariant models involve only three independent FFs at the peak of the resonance, since the virtual background of lower-spin components of the field is not present on the mass shell. As there are three FFs in every model, increasing the degree of symmetry of a model only can select a higher-symmetry part of the FFs and, consequently, increase the degree of the polynomials in $Q^2$ that accompany dispersionlike FF functions in observables.

Comparing Eqs.~\eqref{A32}--\eqref{S12} evaluated in the point- and gauge-invariant model for $\ell = 1$ (a vector-spinor field) with the helicity amplitudes and Jones-Scadron FFs calculated in conventional \cite{Devenish1976, la-06, vereshkov:073007} and gauge-invariant models \cite{pa-07} shows, however, that they are equivalent up to linear redefinition of the Lagrangian FFs $F_f^{(N)}(Q^2)$. In particular, we can obtain the helicity amplitudes in the conventional model \cite{Devenish1976, la-06, vereshkov:073007, pa-07} by the following redefinition:
\begin{align*}
	F_1^{(N)}(Q^2)
	&{}= -\frac12 \left[ G_2^{(N)}(Q^2) + G_3^{(N)}(Q^2) \right],
	\\
	F_2^{(N)}(Q^2)
	&{}= \frac{M_N}{M_R} G_1^{(N)}(Q^2)
	     + \frac12 \left[ G_2^{(N)}(Q^2) - G_3^{(N)}(Q^2) \right],
	\\
	F_3^{(N)}(Q^2) &{}= - G_3^{(N)}(Q^2).
\end{align*}
Likewise, we can obtain the gauge-invariant helicity amplitudes \cite{pa-07} by redefining the FFs as follows
\begin{align*}
	F_1^{(N)}(Q^2) &{}= g_m^{(N)}(Q^2) + g_e^{(N)}(Q^2),
	\\
	F_2^{(N)}(Q^2) &{}= g_m^{(N)}(Q^2) - g_e^{(N)}(Q^2) - 2 g_c^{(N)}(Q^2),
	\\
	F_3^{(N)}(Q^2) &{}= -2 g_c^{(N)}(Q^2).
\end{align*}

Thus, we have come to the conclusion that all three models---the point- and gauge-invariant model being developed, the gauge-invariant model \cite{pa-07}, and conventional one \cite{Devenish1976, la-06, vereshkov:073007}---are equivalent on the mass shell of the resonance up to a linear redefinition of the FFs. It should be noted, however, that these models differ if the resonance is off the mass shell. For example, the one-pion decay of virtual resonance exhibits different $W_R$ dependence in the gauge-invariant model \cite{1998PhRvD..58i6002P} and the point- and gauge-invariant one \eqref{eq:NRpi}:
\begin{multline}
	\langle N\pi \vert \int\mathrm{d}^4 x \frac{if}{m_\pi M_R^2} \bar \Psi^{[\mu\nu]} \Gamma_{[\mu\nu][\lambda\sigma]} \gamma_5 N^{,\lambda} \pi^{,\sigma} \vert R^* \rangle
	\\{}=
	W_R^2 \langle N\pi \vert \int\mathrm{d}^4x \frac{f}{m_\pi M_R} \bar \Psi^{[\mu\nu]} e_{\mu\nu\lambda\sigma} \gamma_5 \gamma^\lambda N \pi^{,\sigma} \vert R^* \rangle,
\end{multline}
where $\Psi_{[\mu\nu]} = \partial_\mu\Psi_\nu - \partial_\nu\Psi_\mu$.

In addition, linear redefinitions could change the physical meaning of the FFs mixing helicity conserving and nonconserving contributions. This could obscure the $Q^2$ dependencies of the ratios of the FFs such as their low-$Q^2$ scaling \cite{PhysRevLett.91.092003, 2009vereshkov}.


\subsection{\label{asymptotes}High-$Q^2$ behavior of transition form factors}

At very high momentum transfer, PQCD predicts the scaling behavior of the photoabsorption amplitudes to be \cite{ca-86, 1988carlson, st-93, 2004idilbi}
\begin{equation}\label{AAS}
\begin{gathered}
	A_{3/2} \sim \frac{1}{Q^5\ln^{n_1}{\displaystyle \frac{Q^2}{\Lambda^2}}},
	\quad
	A_{1/2} \sim \frac{1}{Q^3\ln^{n_2}{\displaystyle \frac{Q^2}{\Lambda^2}}},
	\\
	S_{1/2} \sim \frac{1}{Q^3\ln^{n_3}{\displaystyle \frac{Q^2}{\Lambda^2}}},
	\quad
	n_2-n_1 \approx 2.
\end{gathered}
\end{equation}

Comparing Eqs.~\eqref{AAS} with the point- and gauge-invariant helicity amplitudes \eqref{A32}--\eqref{S12}, we get high-$Q^2$ behavior of the FFs,
\begin{equation}\label{FFF}
	F_{1,3} \sim \frac{1}{Q^{6+2\ell}\ln^{n_{1,3}}{\displaystyle \frac{Q^2}{\Lambda^2}}},
	\quad
	F_2 \sim \frac{1}{Q^{4+2\ell}\ln^{n_2}{\displaystyle \frac{Q^2}{\Lambda^2}}},
\end{equation}
where $n_1 < n_2 < n_3$.

This implies that the FFs $F_1(Q^2) \sim Q^{-2\ell-1} A_{3/2}$, $F_2(Q^2) \sim Q^{-2\ell-1} A_{1/2}$, $F_3(Q^2) \sim Q^{-2\ell-3} S_{1/2}$ acquire (in the asymptotic domain) the statuses of, respectively, the FF of the processes involving flips of two quark helicities, the non-helicity-flip FF, and the helicity-flip FF.

It should be noted that such high-$Q^2$ properties of the FFs are naturally consistent with the classification of the FFs in terms of the differential order of the interaction Lagrangian. Indeed, the first and the third terms of the Lagrangian \eqref{eq:LRNV} involve baryon fields of opposite chiralities and, thus, describe electroproduction with the flip of baryon helicity. The second term, contrarily, links the nucleon and resonance fields of the same chirality and, consequently, amounts to helicity-conserving interactions.


\subsection{Summary of properties of the point- and gauge-invariant interactions}

The point- and gauge-invariant Lagrangians \eqref{eq:NRpi} and \eqref{eq:LRNV} are specific among other consistent gauge-invariant Lagrangians. (i) All terms of the minimally local Lagrangians are defined uniquely by the symmetry. (ii) The point- and gauge-invariant FFs have simple power-logarithmic high-$Q^2$ asymptotes \eqref{FFF}. (iii) The symmetry classifies the $NR$-transition FFs in terms of the differential order of the corresponding Lagrangian vertex. The first and the third FFs describe the interactions with hadron-helicity flips, while the second one is for the helicity-conserving interactions. (iv) This classification is naturally consistent with PQCD interpretation of the FFs in the asymptotic domain. The helicity-(non)conserving amplitudes are proportional to the corresponding helicity-(non)conserving FFs. (v) The tensor-matrix structure of the Lagrangians \eqref{eq:NRpi} and \eqref{eq:LRNV} are unified by the symmetry in regard to the spin of the resonance. All couplings are expressed through the universal matrix $\Gamma_{\bar A \bar B}$. The pre-FF polynomials in the helicity amplitudes \eqref{A32}--\eqref{S12} are the same for any spin of the resonance. Consequently, all the properties (i)--(iv) are valid for arbitrarily high resonance spin.

Therefore, the symmetry of the model (particularly, the point invariance) can be considered as a tool to eliminate ambiguities in choosing the Lagrangian terms and to classify FFs.

It is important to note that the properties (i)--(iv) of the point- and gauge-invariant theory of the nucleon transitions to higher-spin resonances are shared with the well-known theory of the elastic nucleon FFs and the theory of the nucleon transitions to spin-$\frac12$ resonances briefly reviewed further.


\subsubsection{Elastic nucleon interactions}

The elastic $NNV$ vertex \cite{1950rosenbluth, 1960ernst} is determined uniquely by its gauge symmetry,
\begin{align}\label{eq:elastic}
	\mathscr{L} = \sum_V \bar N \left[ g_{1}^{(V)} \gamma_\mu V^\mu
	- \frac{i g_{2}^{(V)}}{2M_N} \sigma_{\mu\nu} V^{\mu\nu} \right] N.
\end{align}
The isotopic indices and matrices are omitted for simplicity.

The first Dirac term comes from the covariant derivative in the kinetic term of the nucleon Lagrangian and the second Pauli term is the only gauge-invariant expression involving just one field derivative. Note also that the Dirac and Pauli terms of the Lagrangian come from the different differential orders of the Lagrangian.

The elastic analogs of the $NR$ helicity amplitudes are Sachs FFs that are related to kinematic-free Lagrangian FFs as follows:
\begin{align}
	G_E(Q^2) &{}= F_1(Q^2) - \frac{Q^2}{2M_N^2} F_2(Q^2),
	\\
	G_M(Q^2) &{}= F_1(Q^2) + F_2(Q^2),
	\\
	F_f(Q^2) &{}= \sum_V \frac{g_f^{(V)} m_V^2}{Q^2+m_V^2}, \qquad f=1,\,2.
\end{align}

The high-$Q^2$ behavior of the FFs is predicted by PQCD \cite{PhysRevD.11.1309, PhysRevLett.91.092003}:
\begin{equation}\label{eq:3e1}
	F_f(Q^2) \sim \frac{1}{Q^{2p_f} \displaystyle \ln^{n_f}{\frac{Q^2}{\Lambda^2}}},
\end{equation}
for $p_1=2$, $n_1 \approx 2$ and $p_2=3$, $n_2 \approx 0$.
As follows from Eq.~\eqref{eq:3e1}, at asymptotically high $Q^2$ the magnetic FF scales like the Dirac one and the electric FF scales like the Pauli one \cite{PhysRevD.11.1309, PhysRevLett.91.092003}:
\begin{equation}
	G_E(Q^2) = - \frac{Q^2}{2M_N^2} F_2(Q^2),
	\qquad
	G_M(Q^2) = F_1(Q^2).
\end{equation}
We can see now that the FFs $F_f(Q^2)$ have distinct interpretations in terms of the underlying quark dynamics---the Dirac FF is the FF of the processes conserving quark helicities, while the Pauli FF is the FF for the processes with quark-helicity flips. This is in accord with the classification of the corresponding Lagrangian terms by their differential order---the Dirac interaction relates nucleons of the same chirality, while the Pauli one relates operators of different chiralities.


\subsubsection{$NR$ interactions ($J_R=\frac12$)}

The interaction Lagrangian for the transitions of nucleons to spin-$\frac12$ resonances is
\begin{align}
	\mathscr{L} &{} = \sum_V \left( \mathscr{L}_1^{(V)} + \mathscr{L}_2^{(V)} \right) + \text{H.c.},
	\\
	\mathscr{L}^{(V)}_1
	&{} = \frac{i g_{2}^{(V)}}{8M_N} \bar R \sigma_{\mu\nu} \gamma_R N V^{\mu\nu},
	\\
	\mathscr{L}^{(V)}_2
	&{} = \frac{g_{1}^{(V)}}{8M_N^2}
	\left( \bar R_{,\mu} \gamma_\nu \gamma_R N + \bar R \gamma_\mu \gamma_R N_{,\nu} \right)
	V^{\mu\nu}.
\end{align}
Here $\mathscr{L}^{(V)}_1$ is the only possible Lagrangian with one field derivative. It describes interactions of baryons with the same chiralities. The second term describes the interactions of baryons with different chiralities.

The helicity amplitudes are
\begin{align}\label{eq:A1212}
	 A_{1/2}^{(p,n)}(Q^2) = {}&\sqrt{2N_0(Q^2)} 
	\notag\\&{}\times
	\left[Q^2 F_1^{(p,n)}(Q^2)+ \mu_\pm M_N F_2^{(p,n)}(Q^2) \right],
\end{align}
\begin{align}\label{eq:S1212}
	 S_{1/2}^{(p,n)}(Q^2) = {}&\pm \frac{Q_+ Q_-}{2M_R} \sqrt{N_0(Q^2)}
	\notag\\ &{} \times
 \left[\mu_\pm F_1^{(p,n)}(Q^2)-M_N F_2^{(p,n)}(Q^2)\right],
\end{align}
where $N_0(Q^2) = \pi\alpha Q^2_\mp /[M_N^5(M_R^2-M_N^2)]$. In Eqs. \eqref{eq:A1212} and \eqref{eq:S1212} the top signs correspond to the resonances of the positive parity like $N(1440)$, while the bottom ones are for the negative-parity resonances like $N(1535)$.

Perturbative QCD predictions \eqref{AAS} for $A_{1/2}(Q^2)$ and $S_{1/2}(Q^2)$ at high $Q^2$ result in the power-logarithmic behavior \eqref{eq:3e1} of the FFs for $p_1 = p_2 = 3$, $n_1-n_2 \approx 2$. The following relations between helicity-(non)flip amplitudes and FFs are valid in the asymptotic region:
\begin{align}
	A_{1/2}(Q^2) \sim Q^3 F_1(Q^2),
	\qquad
	S_{1/2}(Q^2) \sim Q^3 F_2(Q^2).
\end{align}


\section{\label{fit section}point- and gauge-invariant interactions of on-shell resonances}

\subsection{\label{VMD}Transition form factors in a vector-meson--dominance model}

The vector-meson--dominance (VMD) model \cite{2007EPJA...34..223V, vereshkov:073007} which we utilize to fit experimental data assumes that a photon propagating inside nucleon excites all the modes of a hadronic string carrying quantum numbers of the photon $J^{PC}=1^{--}$. Thus, all the observed vector mesons (and, perhaps, hypothetical ones) should be incorporated in the model. The FFs $F^{(p,n)}_f(Q^2)$ are given by the sum over isosinglet and isovector contributions:
\begin{equation}\label{Fvdm}
	F^{(p,n)}_f(Q^2)=
	\frac12 \sum_{k=1}^K \Biggl[
	\frac{\varkappa^{(\omega)}_{f k}(Q^2) m_{(\omega)k}^2}{Q^2+m_{(\omega)k}^2} \pm
	\frac{\varkappa^{(\rho)}_{f k}(Q^2) m_{(\rho)k}^2}{Q^2+m_{(\rho)k}^2} \Biggr],
\end{equation}
where $f=1,\,2,\,3$. Because of the value of the $\Delta$ isospin, $\rho$ mesons are only intermediaries in the $N\Delta$ coupling, i.e., $\varkappa_{f k}^{(\omega)}(Q^2)=0$ for $\Delta$ resonances.

In Eq.~\eqref{Fvdm} it is supposed that the meson spectrum is truncated at highly excited broad-width states that cannot be reliably separated from the continuum. In fitting experimental data (see Sec.~\ref{fit}), however, we include only five $\rho\omega$ families in the model, because this is enough to make the VMD model agree with all the experimental data at spacelike momentum transfers and to attain correct high-$Q^2$ behavior predicted by perturbative QCD \eqref{AAS}. We believe that these mesons give major contributions to the observables, although to prove this statement one needs abundant data on the nucleon transition FFs at the timelike momentum transfers.

Since our model is designed for a global (in $Q^2$) fit of transition FFs, we should guarantee correct high-$Q^2$ behavior \eqref{FFF}. This raises a question of how to comply consistently with the requirement \eqref{FFF}. There are two physical considerations that can be invoked to resolve the problem in the framework of effective field theory.

First of all, since baryons are composite particles, the effects of nonlocality may manifest themselves for $Q^2$ increasing inverse nucleon size squared $R_N^{-2} = (0.2 \text{ GeV})^2$. We can therefore consider transition FFs $F_f^{(p,n)}(Q^2)$ as nonlocal and use the Taylor expansion of nonlocal FFs \eqref{FF series} to suppress growing polynomial functions in the helicity amplitudes \eqref{A32}--\eqref{S12}.

Second, to retain correct high-$Q^2$ behavior of the FFs is also possible in minimally local effective field theory. To this end, one should incorporate the higher excitations of the vector mesons in the model and impose superconvergence relations on the parameters of the meson spectrum \cite{du-03, 2007EPJA...34..223V}.

What option of the two ones above should be preferred is an open issue. In what follows we try to stick to the second one, which is simpler and allows us not to transcend the framework of local field theory.

To assure correct high-$Q^2$ behavior \eqref{AAS} of the dispersionlike expansions of the FFs \eqref{Fvdm}, we assume the following.
(i) The $Q^2$-dependence of the expansion coefficients is independent of the meson-family index $k$:
\begin{align}
\varkappa_{kf}(Q^2) = \frac{\varkappa_{kf}(0)}{L_f(Q^2)}.
\end{align}
(ii) The logarithmic corrections in Eq.~\eqref{AAS} are taken into account by phenomenological interpolation functions:
\begin{align}\label{logs}
L_f(Q^2) = \left[1+b_f\ln\left(1+\frac{Q^2}{\Lambda^2} \right)+
a_f\ln^2\left(1+\frac{Q^2}{\Lambda^2}\right)\right]^{n_f/2}.
\end{align}
The logarithmic interpolation functions \eqref{logs} take account of short-distance quark-gluon processes influencing the photon transition to mesons inside nucleon, i.e., at $Q^2 > R_N^{-2} = (0.2 \text{ GeV})^2$.
\newline
(iii) Finally, after applying asymptotic restrictions \eqref{FFF} for dispersionlike expansions
\begin{equation}\label{15}
	\sum_{k=1}^{K} \frac{m_k^2 \varkappa_{kf}(0)}{m_k^2+Q^2}
	 = -\sum_{i=1}^\infty \left( -\frac{1}{Q^2} \right)^i \sum_{k=1}^{K}{m_k^{2i}\varkappa_{kf}(0)}
\end{equation}
we have superconvergence relations
\begin{equation}\label{SR}
	\sum_{k=1}^K m_{(\omega,\,\rho)k}^{2n} \varkappa_{kf}^{(\omega,\,\rho)}(0) =0,
\end{equation}
where $n=2,\,3$ for $\ell = J_R - \frac12 = 0$; $n = 2,\,3,\dots\,4+ \ell$ for $f=1,\,3$, $\ell \geqslant 1$; and $n = 2,\dots\,3+ \ell$ for $f=2$, $\ell \geqslant 1$. As we can see, the minimal number of the mesons enough to saturate the superconvergence relations increases with the spin of the resonance, $K_\text{min} = 3+\ell$ for $\ell = 0,\,1,\dots$.

In what follows it is convenient to normalize coefficients $\varkappa_{kf}^{(\omega,\,\rho)}(0)$ by low-energy constants $F_f^{(\rho, \omega)}(0) = \sum_{k=1}^K \varkappa^{(\rho)}_{fk}(0)$:
\begin{align}
	h_{kf}^{(\omega,\,\rho)} = \frac{\varkappa_{kf}^{(\omega,\,\rho)}(0)}{F_f^{(\rho, \omega)}(0)}.
\end{align}
The parameters $h_{kf}^{(\omega,\,\rho)}$ satisfy the superconvergence relations \eqref{SR} as the parameters $\varkappa_{kf}^{(\omega,\,\rho)}(0)$ do. In addition, the following sum rule is valid for all $f=1,\,2,\,3$:
\begin{align}\label{eq:sr}
	\sum_{k=1}^{K} h_{kf}^{(\omega,\,\rho)} = 1.
\end{align}


\subsection{\label{fit}Data analysis}


\subsubsection{Resonance $\Delta(1232)$}

Since the resonance $\Delta(1232)$ carries the isospin $3/2$, it couples only to the isovector $\rho$ mesons. Therefore, in the VMD model described, the FFs for the $N\Delta(1232)$ transition have the form
\begin{align}\label{F1F2F3}
	F_f^{(p)}(Q^2)
	 = \frac{F_f^{(p)}(0)}{L_f(Q^2)}
	   \sum_{k=1}^K \frac{h^{(\rho)}_{fk} m^2_{(\rho)k}}{Q^2+m^2_{(\rho)k}},
	\quad f=1,\,2,\,3,
\end{align}
where $m^2_{(\rho)k}$ are the masses of $\rho$ mesons listed in Table~\ref{tab:mesons}, the couplings $h^{(\rho)}_{fk}$ satisfy the superconvergence relations \eqref{SR} and the sum rule \eqref{eq:sr}, and logarithmic interpolation functions $L_f(Q^2)$ are given by Eq.~\eqref{logs}.

\begin{table}
\caption{\label{tab:mesons}Vector-meson masses \cite{PDG}. The isosinglet mesons $\omega(1960)$ and $\omega(2205)$ are listed in the section ``Further states'' of Ref. \cite{PDG}. The last column gives an averaged mass $m_k = \left[ (m_{(\omega)k}^2 + m_{(\rho)k}^2)/2 \right]^{1/2}$.}
\small
\begin{ruledtabular}
\begin{tabular}{cccccc}
$k$&$$&$m_{(\rho)k}$ (GeV) &&$m_{(\omega)k}$ (GeV)&$m_k$ (GeV)\\
\hline
1 & $\rho(770)$  & 0.77549 & $\omega(782)$  & 0.78265 & 0.77908 \\
2 & $\rho(1450)$ & 1.465   & $\omega(1420)$ & 1.425   & 1.445   \\
3 & $\rho(1700)$ & 1.720   & $\omega(1650)$ & 1.670   & 1.695   \\
4 & $\rho(1900)$ & 1.885   & $\omega(1960)$ & 1.960   & 1.923   \\
5 & $\rho(2150)$ & 2.149   & $\omega(2205)$ & 2.205   & 2.177
\end{tabular}
\end{ruledtabular}
\end{table}

The data on the $Q^2$ dependence of the $N\Delta(1232)$ transition \cite{PDG, fr-99, sp-05, sp-08, 2009aznauryan, ka-01, 2008stave, 2009villano, 2007Julia-Diaz, el-06, ke-05} were fitted in the model that involves five lightest $\rho$ mesons. Hence, only four of the parameters $h^{(\rho)}_{fk}$ are independent, while other eleven are calculated from Eqs.~\eqref{SR}--\eqref{eq:sr}. Also six small free parameters $a_f$ and $b_f$ are introduced by the functions $L_f(Q^2)$ to comply with the logarithmic corrections to the high-$Q^2$ scaling \eqref{FFF}.

The free parameters are restricted so that ratios of the FFs do not deviate from a perturbative scaling limit by more than 0.1\% at $Q^2 \geqslant 0.4 \text{ GeV}^2$:
\begin{equation}\label{FFR scaling}
	\frac{F_f(Q^2)}{F_2(Q^2)} \propto \frac{1}{Q^2} \left( \ln \frac{Q^2}{\Lambda^2} \right)^{n_2-n_f}, \quad f = 1,\; 3.
\end{equation}

\begin{figure}
	\center\includegraphics[width=0.8\linewidth]{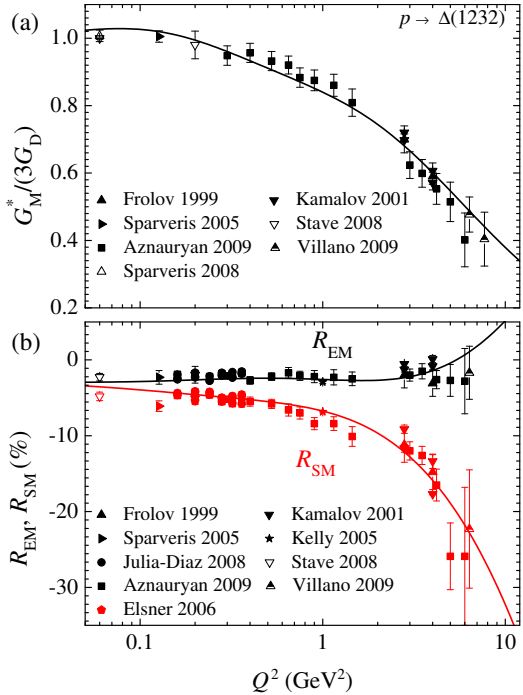}
	\caption{\label{fig:Delta(1232)}(Color online) Magnetic FF $G_\text{M}^*(Q^2)$ normalized by $G_\text{D}(Q^2) = (1+Q^2/0.71)^{-2}$ and the amplitude ratios $R_\text{EM}(Q^2)$ and $R_\text{SM}(Q^2)$ for the transition $p\gamma^*\to \Delta(1232)$ ($\chi^2/\text{DOF} = 1.51$). The data points are as follows: Frolov 1999 \cite{fr-99}, Sparveris 2005 \cite{sp-05}, Sparveris 2008 \cite{sp-08}, Aznauryan 2009 \cite{2009aznauryan}, Kamalov 2001 \cite{ka-01}, Stave 2008 \cite{2008stave}, Villano 2009 \cite{2009villano}, Juli{\'a}-D{\'{\i}}az 2008 \cite{2007Julia-Diaz}, Elsner 2006 \cite{el-06}, Kelly 2005 \cite{ke-05}.}
\end{figure}

The overall quality of the fit by Eqs.~\eqref{A32}--\eqref{S12} and \eqref{F1F2F3} is $\chi^2/\text{DOF} = 1.71$. The adjusted parameters are set out in Table~\ref{tab:param32}. The corresponding curves are depicted in Fig.~\ref{fig:Delta(1232)}. The magnetic dipole FF is defined in the Jones-Scadron convention \cite{jo-sc}:
\begin{align*}
	G_\text{M}^* &(Q^2) =
	\\
	&- \left[ \frac{M_N^3 (M_\Delta^2-M_N^2)}{2 \pi \alpha (M_\Delta+M_N)^2} \right]^{1/2} \frac{A_{1/2} + \sqrt{3} A_{3/2}}{\left[ Q^2+(M_\Delta-M_N)^2 \right]^{1/2}}.
\end{align*}
The ratios $R_\text{EM}$ and $R_\text{SM}$ of the electric and Coulomb quadrupole moments to the magnetic dipole one are written as
\begin{align}
	R_\text{EM} = \frac{A_{1/2}-\displaystyle \frac{1}{\sqrt{3}} A_{3/2}}{A_{1/2}+ \sqrt{3} A_{3/2}},
	\quad
	R_\text{SM} = \frac{\sqrt{2} S_{1/2}}{A_{1/2}+ \sqrt{3} A_{3/2}}.
\end{align}

\begin{figure}
	\center\includegraphics[width=0.8\linewidth]{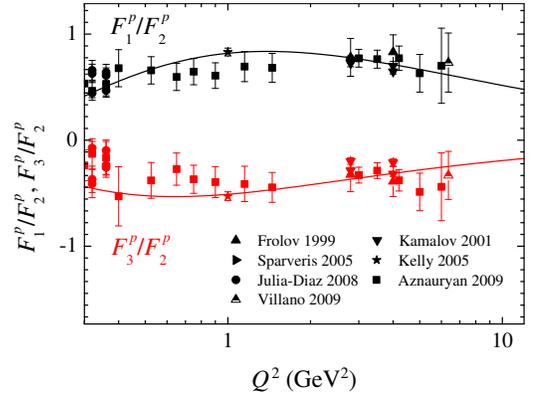}
	\caption{\label{fig:FFRs}(Color online) FF ratios extracted using Eqs.~\eqref{A32}--\eqref{S12} from the data on the ratios $R_\text{EM}$ and $R_\text{SM}$. The fit curves agree with a low-energy scaling \eqref{FFR scaling} at $Q^2 = 0.4$--7 GeV$^2$. The references to the experimental data are the same as for Fig.~\ref{fig:Delta(1232)}.}
\end{figure}

The ratios of the FFs $F_{1,3}/F_2$ extracted from the available experimental data \cite{fr-99, ka-01, PDG, sp-05, ke-05, 2007Julia-Diaz, 2008stave, 2009aznauryan, 2009villano} on $R_\text{EM}$ and $R_\text{SM}$ are depicted in Fig.~\ref{fig:FFRs}. The agreement of the scaling hypothesis \eqref{FFR scaling} with the data for $Q^2 \geqslant 0.4 \text{ GeV}^2$ is at the level of $\chi^2/\text{DOF} = 1.03$. The good agreement testifies that the hypothesis of the low-$Q^2$ scaling of the FF ratios is adequate to describe the $Q^2$-evolution of the ratios $R_\text{EM}$ and $R_\text{SM}$ for the $N\Delta(1232)$ transition.

\begin{table*}
\caption{\label{tab:param32}Fit parameters.}
\begin{ruledtabular}
\begin{tabular}{lD{.}{.}{-1}D{.}{.}{-1}D{.}{.}{-1}D{.}{.}{-1}D{.}{.}{-1}}
&\multicolumn{1}{r}{$\Delta(1232)$}&\multicolumn{1}{r}{$N(1440)$}&\multicolumn{1}{r}{$N(1520)$}&\multicolumn{1}{r}{$N(1535)$}&\multicolumn{1}{r}{$N(1680)$} \\
$\chi^2/\text{DOF}$ & 1.71 & 0.97 &0.87& 0.65 & 1.05 \\
\hline
$F^{(p)}_1(0)$ &  0.4203 &  0.0782 & 0.6899 & 0.3406 & 0.2458 \\
$F^{(p)}_2(0)$ &  0.7556 & -0.2490 & 1.4932 & 0.2574 & 0.4643 \\
$F^{(p)}_3(0)$ & -0.3185 &   \tabd & 0.5602 &  \tabd & 0.0855 \\

$h^{(p)}_{14}$ &   \tabd & -0.0437 &  \tabd & -0.9968 &  \tabd \\
$h^{(p)}_{15}$ &  0.0001 &   \tabd &  \tabd &   \tabd &  \tabd \\
$h^{(p)}_{24}$ & -0.8642 &  1.5790 & 0.1685 & -2.4420 &  \tabd \\
$h^{(p)}_{25}$ &  0.8683 &   \tabd &  \tabd &   \tabd & 0.1383 \\
$h^{(p)}_{35}$ & -0.0341 &   \tabd &  \tabd &   \tabd &  \tabd \\

$a_1$ &  0.0090 & -0.4820 & 0.0011 & -0.2379 &  0.6242 \\
$b_1$ & -0.1413 &  0.0754 & \tabd  &  0.0242 & -1.0080 \\
$a_2$ &  0.2268 & -0.5338 & 0.0032 & -0.6667 &  0.0389 \\
$b_2$ & -0.1339 &  0.1228 & \tabd  &  0.2355 & -0.1034 \\
$a_3$ &  0.9416 &  \tabd  & 0.0012 & \tabd   &  0.0883 \\
$b_3$ & -0.0377 &  \tabd  & \tabd  & \tabd   & -0.5403 \\

$\Lambda$ (GeV)& 0.2950 & 0.3 & 0.101 & 0.3 & 0.3 \\
\end{tabular}
\end{ruledtabular}
\end{table*}


\subsubsection{Resonances $N(1440)$, $N(1520)$, $N(1535)$, $N(1680)$}

Though both $\rho$ and $\omega$ mesons contribute to the excitation of the nucleon resonances, currently there are no measurements of neutron helicity amplitudes, except for the photoproduction data~\cite{PDG}. Thus, in the adopted VMD model, it is hardly possible to distinguish reliably isovector and isoscalar contributions to the FFs. Because of this reason, in the following we neglect singlet-triplet mass splitting and suppose that $\rho$ and $\omega$ mesons propagate in the nucleon medium identically; i.e., $L^{(\rho)}_f(Q^2) \equiv L^{(\omega)}_f(Q^2)$. In such a model, proton transition FFs depend on half as many independent parameters as the FFs \eqref{Fvdm} and are written as
\begin{align}\label{F123p}
	F_f^{(p)}(Q^2)
	 = \frac{F_f^{(p)}(0)}{L_f(Q^2)}
	   \sum_{k=1}^K \frac{h^{(p)}_{fk} m^2_{k}}{Q^2+m^2_{k}}, \quad f=1,\,2,\,3,
\end{align}
with $m_k^2 = (m_{(\omega)k}^2 + m_{(\rho)k}^2)/2$ being vector meson masses averaged for each singlet-triplet family (see Table~\ref{tab:mesons}). In spite of the simplifications described above, the FFs \eqref{F123p} provide a good fit of the existing data on the helicity amplitudes for the transitions to the resonances $N(1440)$, $N(1520)$, $N(1535)$, and $N(1680)$.

\begin{figure}
	\center\includegraphics[width=0.8\linewidth]{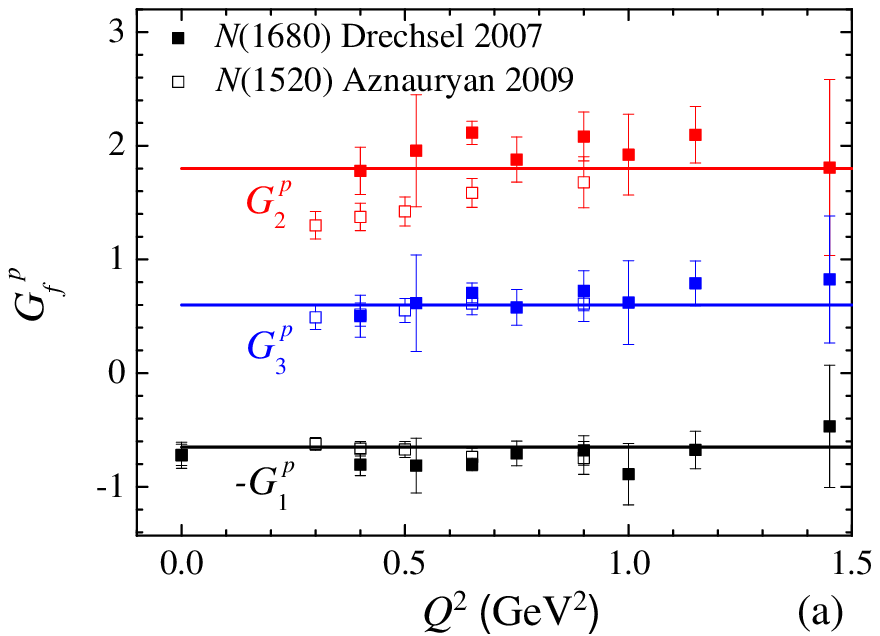}
	\center\includegraphics[width=0.8\linewidth]{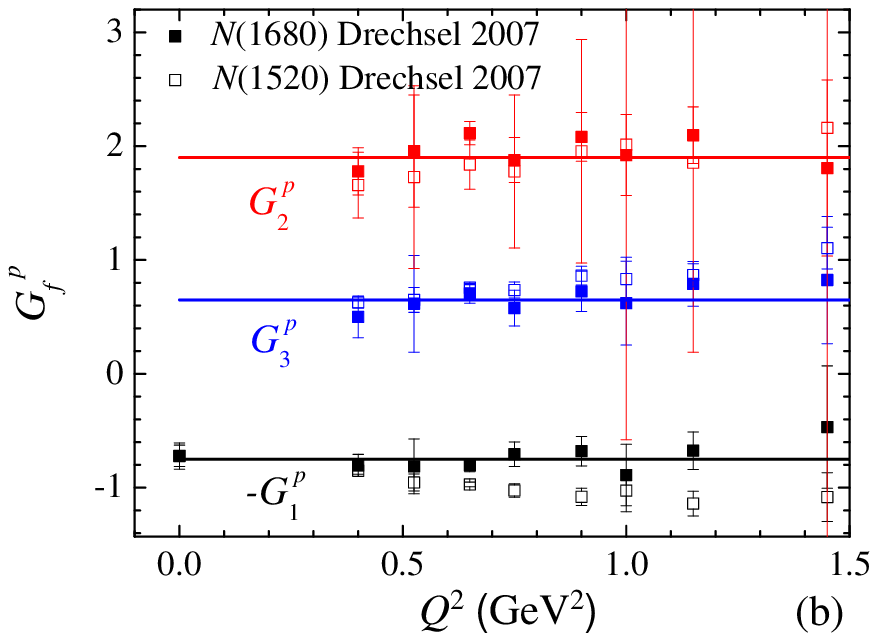}
	\caption{\label{fig:1680vs1520}(Color online) The FFs defined by Eqs.~\eqref{eq:g1520} and \eqref{eq:g1680}. The data points are $N(1520)$ Aznauryan 2009 \cite{2009aznauryan} and $N(1520)$, $N(1680)$ Drechsel 2007 \cite{2007EPJA...34...69D}.}
\end{figure}

We should point out an empirical feature of the experimental data. As we have seen earlier, the relations \eqref{A32}--\eqref{S12} between FFs $F_f(Q^2)$ the helicity amplitudes are universal for resonances with spin parities $J_R = \frac32^-,\ \frac52^+, \dots$, except for the resonance masses and common $Q^2$-dependent factor $(Q_+Q_-)^\ell$. Therefore, it could be interesting to compare FFs for the resonances $N(1520)$ ($J_R = \frac32^-$) and $N(1680)$ ($J_R = \frac52^+$):
\begin{align}
	G_f(Q^2,\ 1520) &= \frac{F_f^{(p)}(Q^2,\ 1520)}{G(Q^2)},\label{eq:g1520}
	\\ \label{eq:g1680}
	G_f(Q^2,\ 1680) &= 1.4 \frac{Q_+Q_-}{M_N^2} \frac{F_f^{(p)}(Q^2,\ 1680)}{G(Q^2)},
\end{align}
where the normalizing function is
\begin{gather}
	G(Q^2) = \prod_{k=1}^4 \left(1 + \frac{Q^2}{m_{V(k)}^2} \right)^{-1}.
\end{gather}
The experimental data for the quantities $G_f(Q^2,\ M_R)$ at $Q^2 < 1.5 \text{ GeV}^2$ are shown in Fig.~\ref{fig:1680vs1520}. It is quite interesting that at these low-momentum transfers these quantities for the transitions $N(1520)$ and $N(1680)$ are mostly constant and very close or even coincide, $G_f(Q^2,\ 1520) \approx G_f(Q^2,\ 1680)$, $f=1,\ 2,\ 3$.

\begin{figure*}
	\center
	\includegraphics[width=0.4\linewidth]{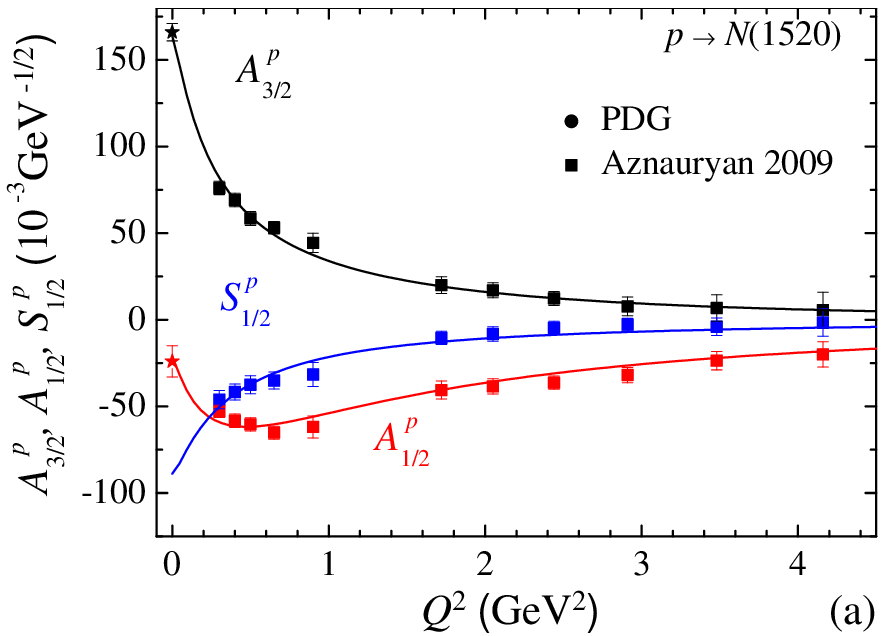}
	\hspace{1cm}\includegraphics[width=0.4\linewidth]{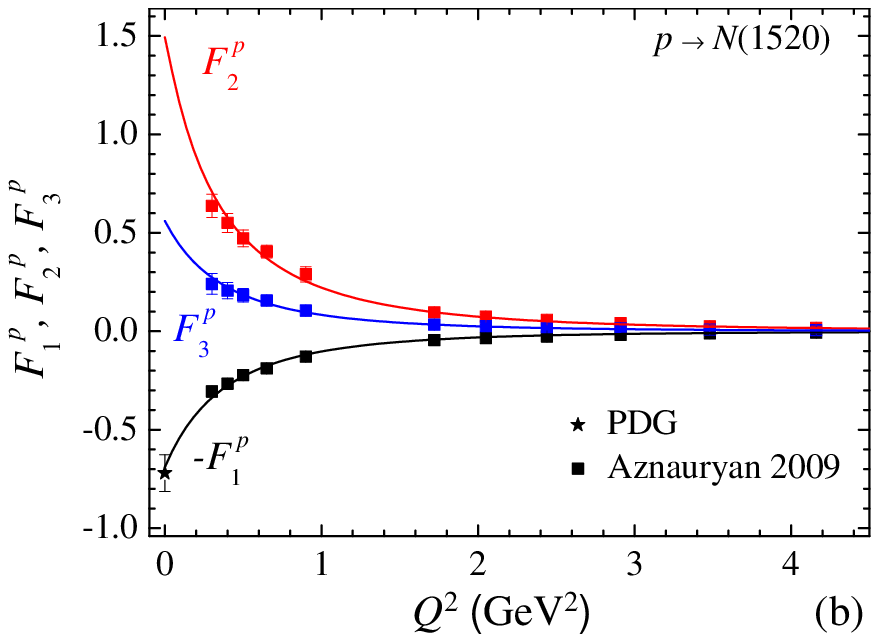}
	\center\includegraphics[width=0.4\linewidth]{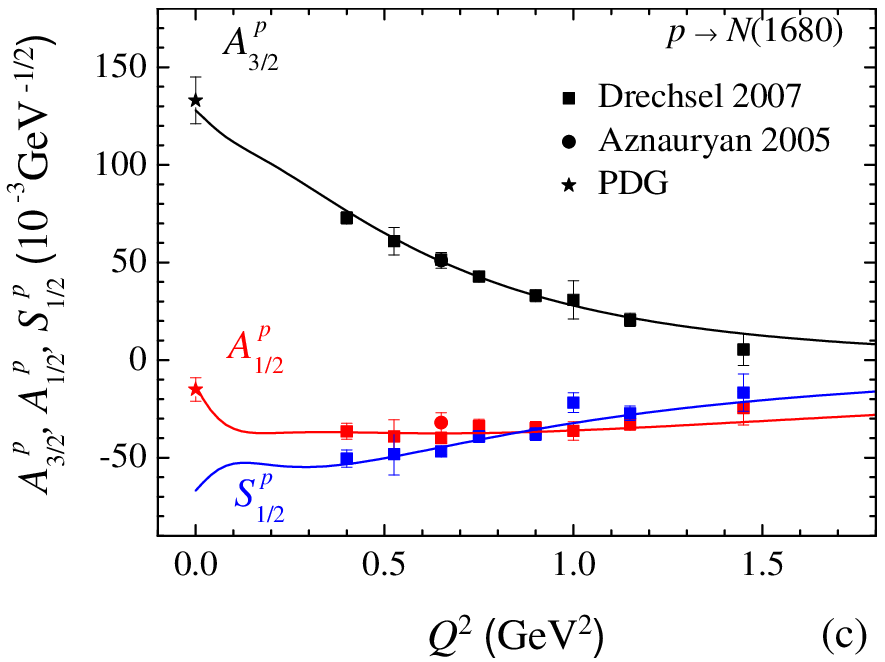}
	\hspace{1cm}\includegraphics[width=0.4\linewidth]{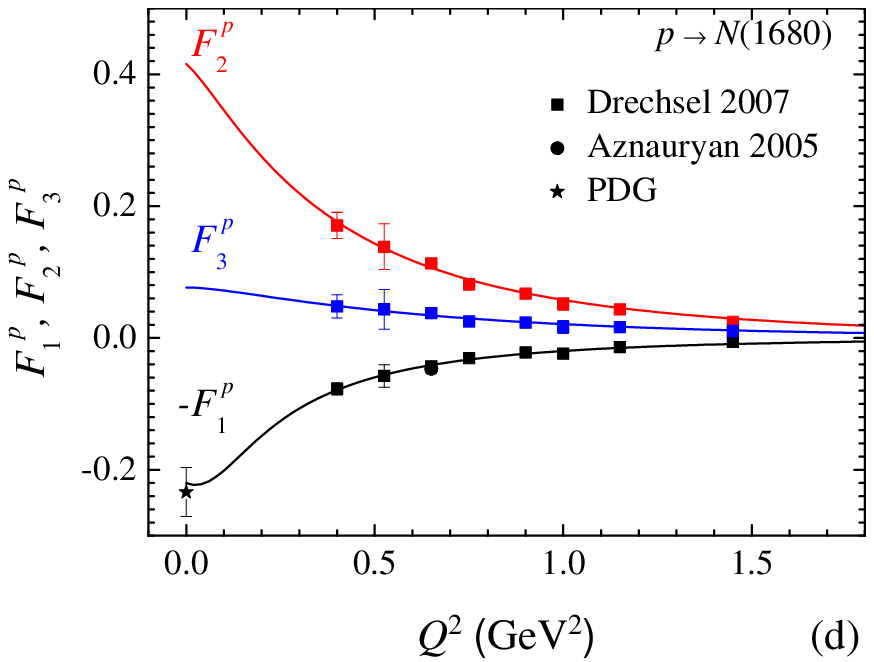}
	\center\includegraphics[width=0.4\linewidth]{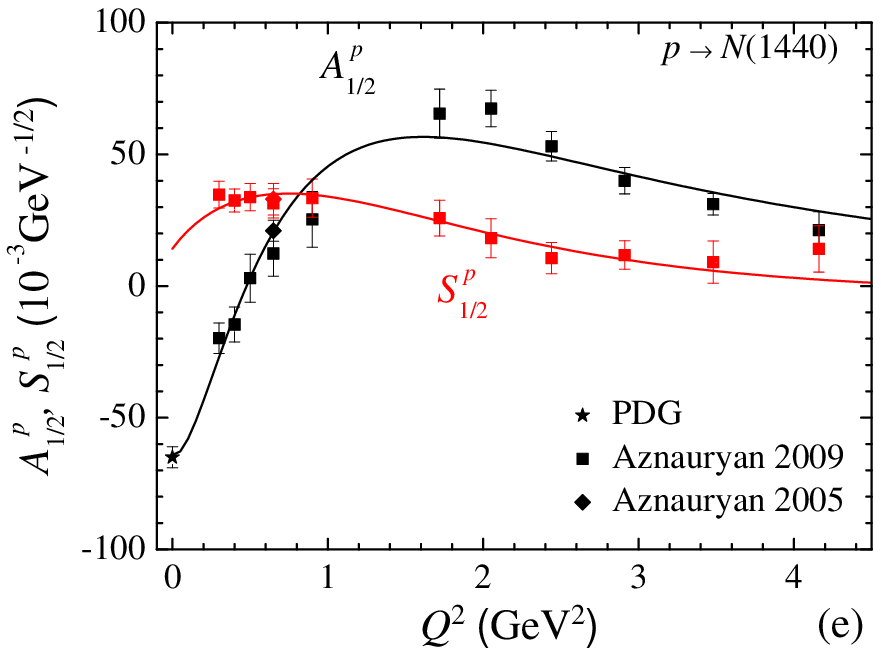}
	\hspace{1cm}\includegraphics[width=0.4\linewidth]{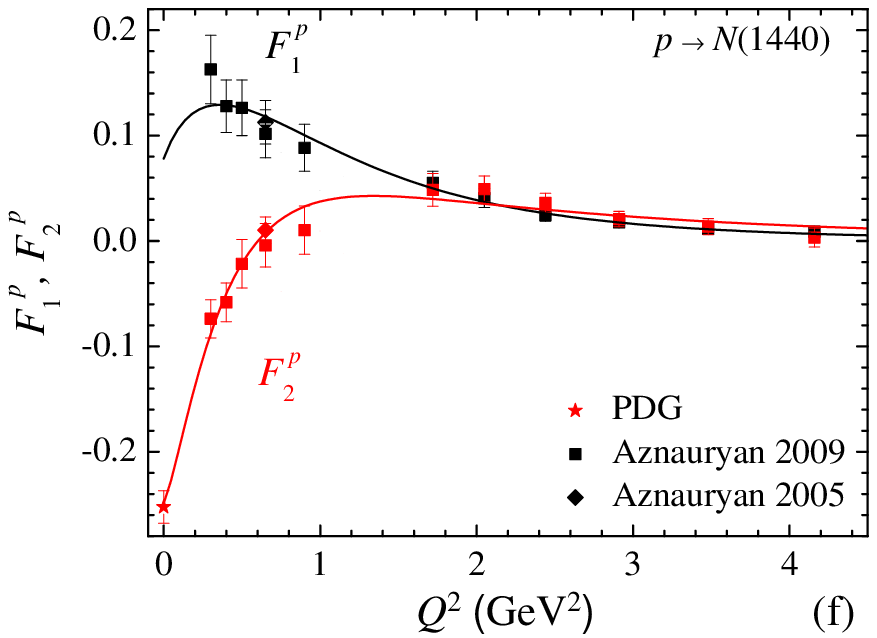}
	\center\includegraphics[width=0.4\linewidth]{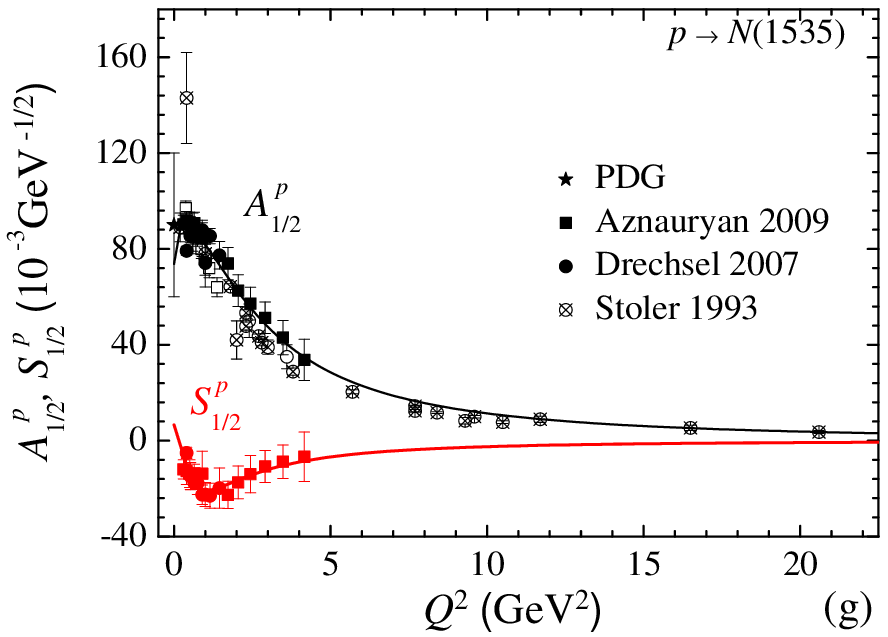}
	\hspace{1cm}\includegraphics[width=0.4\linewidth]{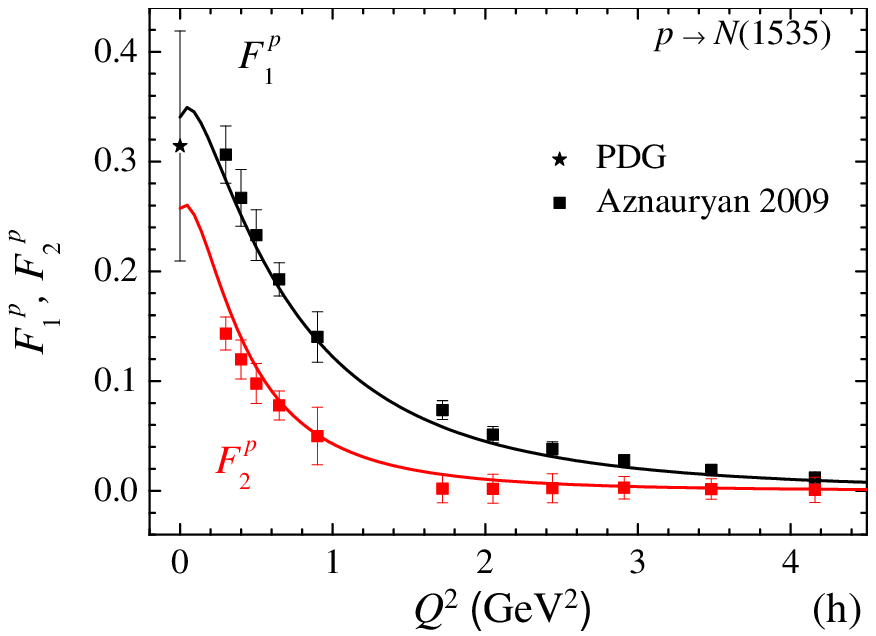}
	\caption{\label{fig:Ns}(Color online) Helicity amplitudes and point- and gauge-invariant FFs for the transitions $p\gamma^*\to N^*_+(1520)$ ($\chi^2/\text{DOF} = 1.05$), $p\gamma^*\to N^*_+(1680)$ ($\chi^2/\text{DOF} = 0.87$), $p\gamma^*\to N^*_+(1440)$ ($\chi^2/\text{DOF} = 0.97$), and $p\gamma^*\to N^*_+(1535)$ ($\chi^2/\text{DOF} = 0.65$). The references to the experimental data are Particle Data Group (PDG) \cite{PDG}, Aznauryan 2005 \cite{az-05b}, Aznauryan 2009 \cite{2009aznauryan}, Drechsel 2007 \cite{2007EPJA...34...69D}, Stoler 1993 \cite{st-93}.}
\end{figure*}

\section{Conclusion}

We have investigated interactions of higher-spin baryon resonances that are invariant under both point and gauge transformations of the RS field. We have discussed some of theoretically appealing properties of such interactions. In particular, it has been shown that such interactions preserve the supplementary conditions of the original RS formalism \eqref{eq:constraints} and consequently involve the correct number of DOF. Another advantage of the point- and gauge-invariant interactions is that they explicitly do not depend on off-shell parameters, either fixed or arbitrary, which is in accord with the redundancy of these parameters in effective field theory \cite{2010PhLB..683..222K}.

We have explicitly written the minimally local point- and gauge-invariant Lagrangian for nucleon-resonance interactions with photons and vector mesons. To this end, we have defined the basis set of $\gamma$-traceless tensor matrices. With respect to a spin of the resonance, the point- and gauge-invariant Lagrangian is unified in its form and properties. (i) All three vertices of the minimally local Lagrangian are constructed with the same tensor matrix $\Gamma_{\bar A \bar B}$, and the tensor-matrix structure of the vertices is the same for arbitrarily high spin of the resonance. (ii) The symmetry of the model classifies vertices as helicity-conserving and nonconserving. (iii) At asymptotically high $Q^2$, the classification is naturally consistent with PQCD predictions leading to simple power-logarithmic scaling behavior of the FFs.

We have fitted the data extracted from resonant electroproduction off nucleon. For this purpose, the Lagrangian FFs were modeled as dispersionlike expansions with poles at vector meson masses. The correct high-$Q^2$ behavior was retained by applying linear superconvergence relations on the parameters of the meson spectrum. The model is in good agreement with the experimental data. It should be noted that good fits to the experimental data on helicity amplitudes of five nucleon and $\Delta$ resonances are obtained in the unified approach, based on five-pole dispersionlike FF expansions \eqref{F1F2F3} satisfying the superconvergence relations. This is an evidence for validity of the VMD model in physics of nucleon excitations.

Based on the available experimental data, we have observed empirically that the point- and gauge-invariant FFs exhibit peculiar $Q^2$ dependencies. The available data set on the transition $N\Delta(1232)$ does not contradict the hypothesis of low-$Q^2$ scaling of the FF ratios, while the FFs for the $NN(1680)$ transition are proportional to the FFs for the $NN(1520)$ transition up to a factor $Q_+ Q_-$.


\appendix

\section{\label{app:a}$\gamma$-traceless matrices}

In this appendix we obtain all possible coupling matrices of the point- and gauge-invariant Lagrangians for baryon resonances with spin $J_R = \ell + \frac12$. To this end, we should consider matrices $\Gamma_{\bar{A}\bar\alpha}$, where $\bar\alpha$ is an arbitrary multi-index and $\bar A = ([\mu_1\nu_1]\dots[\mu_\ell\nu_\ell])$ is a multi-index that is contracted with indices of a higher-spin field strength guaranteeing the gauge invariance of the Lagrangian. The point invariance means that the matrix is $\gamma$ traceless:
\begin{equation}\label{eq:A1}
	\gamma^{\mu_1} \Gamma_{\bar{A}\bar\alpha} = 0.
\end{equation}
As any $4 \times 4$ matrix, the coupling $\Gamma_{\bar{A}\bar\alpha}$ can be written as a linear combination of the identity matrix and the matrices $i\gamma_5$, $\gamma_\mu$, $i\gamma_\mu\gamma_5$, $i\sigma_{\mu\nu}$. Such a combination should be constrained by the requirement \eqref{eq:A1}, which gives all possible $\gamma$-traceless matrices. In the next section this is done straightforwardly for the simplest case of $\ell = 1$. In Sec.~\ref{app:a2} of this appendix the result for $\ell =1 $ is generalized by induction to higher $\ell \geqslant 2$. In Sec.~\ref{app:a3} we prove some important algebraic properties of the matrices considered.


\subsection{$\gamma$-traceless matrices for $\ell = 1$}


\subsubsection{\label{odd rank}Matrices of odd tensor ranks}

Any odd-rank tensor matrix $\Gamma_{[\mu\nu]\bar\alpha}$ can be decomposed as follows:
\begin{align}\label{odd-rank decomposition}
	\Gamma_{[\mu\nu]\bar\alpha}
	 = \gamma^\eta \mathsf{S}_{\eta[\mu\nu]\bar\alpha} + i \gamma^\eta \gamma_5 \mathsf{P}_{\eta[\mu\nu]\bar\alpha},
\end{align}
where the tensor coefficients $\mathsf{S}_{\eta[\mu\nu]\bar\alpha}$ and $\mathsf{P}_{\eta[\mu\nu]\bar\alpha}$ are antisymmetric under interchange of $\mu$ and $\nu$, and defined as
\begin{align}
	\mathsf{S}_{\eta[\mu\nu]\bar\alpha} = -\mathsf{S}_{\eta[\nu\mu]\bar\alpha}
	&{} = \frac14 \Tr \bigl( \gamma_\eta \Gamma_{[\mu\nu]\bar\alpha} \bigr),\label{def of S}
	\\
	\mathsf{P}_{\eta[\mu\nu]\bar\alpha} = -\mathsf{P}_{\eta[\nu\mu]\bar\alpha}
	&{} = \frac14 \Tr \bigl( i\gamma_\eta \gamma_5 \Gamma_{[\mu\nu]\bar\alpha} \bigr).
\end{align}

Requiring the decomposition \eqref{odd-rank decomposition} to obey Eq.~\eqref{eq:A1} we get
\begin{align}
	\gamma^\mu \Gamma_{[\mu\nu]\bar\alpha}
	 = {}&\mathsf{S}^\mu{}_{[\mu\nu]\bar\alpha} + i\gamma_5 \mathsf{P}^\mu{}_{[\mu\nu]\bar\alpha}
	   \notag\\
	   &{}+ \sigma^{\mu\eta} \left( \mathsf{S}_{\eta[\mu\nu]\bar\alpha} +
	    \frac12 e_{\mu\eta}{}^{\omega\rho} \mathsf{P}_{\rho\omega\nu\bar\alpha} \right) =0,\label{g*odd-G}
\end{align}
where the following identity is used
\begin{equation}\label{eq:A6}
	\sigma_{\mu\nu} \gamma_5 = -\frac12 i e_{\mu\nu\lambda\sigma} \sigma^{\lambda\sigma}.
\end{equation}
Granting the orthonormality of the basis set, it follows from Eq.~\eqref{g*odd-G} that
\begin{align}
	\mathsf{S}^\mu{}_{[\mu\nu]\bar\alpha} = 0,\label{S tr-less}
	\\
	\mathsf{P}^\mu{}_{[\mu\nu]\bar\alpha} = 0,\label{P tr-less}
	\\
	\mathsf{S}_{\eta[\mu\nu]\bar\alpha} - \mathsf{S}_{\mu[\eta\nu]\bar\alpha}
	   + e_{\mu\eta}{}^{\omega\rho} \mathsf{P}_{\rho[\omega\nu]\bar\alpha} = 0.\label{coeff sigma}
\end{align}
If we antisymmetrize Eq.~\eqref{coeff sigma} in the indexes $\mu$ and $\nu$, we obtain the simple relation among the tensor coefficients of the decomposition \eqref{odd-rank decomposition},
\begin{align}\label{S(P)}
	\mathsf{S}_{\eta[\mu\nu]\bar\alpha} = \frac12 \bigl( e_{\nu\eta}{}^{\omega\rho} \mathsf{P}_{\rho[\omega\mu]\bar\alpha} - e_{\mu\eta}{}^{\omega\rho} \mathsf{P}_{\rho[\omega\nu]\bar\alpha} - e_{\mu\nu}{}^{\omega\rho} \mathsf{P}_{\rho[\omega\eta]\bar\alpha} \bigr).
\end{align}
Equations \eqref{S tr-less} and \eqref{S(P)} yield another tensor identity for the tensor coefficients $\mathsf{P}_{\eta[\mu\nu]\bar\alpha}$:
\begin{align}\label{eP identity}
	e^{\eta\mu\nu\rho} \mathsf{P}_{\eta[\mu\nu]\bar\alpha} = 0.
\end{align}
Substituting Eq.~\eqref{S(P)} into the decomposition \eqref{odd-rank decomposition} and rearranging it, we get the following expression:
\begin{align}
	\Gamma_{[\mu\nu]\bar\alpha}
	 = \frac12 \biggl[ & \frac12 \gamma^\eta \bigl( e_{\nu\eta\lambda\rho} g_{\mu\sigma}
	                                                - e_{\mu\eta\lambda\rho} g_{\nu\sigma}
	                                                - e_{\mu\nu\lambda\rho} g_{\eta\sigma}
	  \notag\\&{}
	                                                - e_{\nu\eta\sigma\rho} g_{\mu\lambda}
	                                                + e_{\mu\eta\sigma\rho} g_{\nu\lambda}
	                                                + e_{\mu\nu\sigma\rho} g_{\eta\lambda} \bigr)
	  \notag\\&{}
	  +i \gamma^\eta \gamma_5 \bigl( g_{\mu\lambda} g_{\nu\sigma}
	                               - g_{\nu\lambda} g_{\mu\sigma} \bigr) g_{\eta\rho} \biggr]
	   \mathsf{P}^{\rho[\lambda\sigma]}{}_{\bar\alpha}.
	\label{G(P)}
\end{align}
This is the most general decomposition of the odd-rank $\gamma$-traceless and antisymmetric in $\mu\nu$ tensor matrix $\Gamma_{[\mu\nu]\bar\alpha}$. It is seen in the above equation that the tensor matrix $\Gamma_{[\mu\nu]\bar\alpha}$ factorizes into a constant matrix and tensor coefficients $\mathsf{P}_{\rho[\lambda\sigma]\bar\alpha}$ that are antisymmetric in $\lambda\sigma$ and subjected to the conditions \eqref{P tr-less} and \eqref{eP identity}. The matrix in square brackets in Eq.~\eqref{G(P)} does not satisfy the requirement of $\gamma$-tracelessness of itself [Despite the total expression \eqref{G(P)} does, owing to the tensor identities \eqref{P tr-less} and \eqref{eP identity}.] It is possible, however, to make it manifestly $\gamma$ traceless by adding to the right-hand side of Eq.~\eqref{G(P)} the following matrix:
\begin{align}\begin{split}\label{odd rank nill}
	\frac1{12} \gamma^\eta \Bigl[
	 & 4 e_{\lambda\sigma\rho\mu} g_{\nu\eta} - 4 e_{\lambda\sigma\rho\nu} g_{\mu\eta}
	 - e_{\eta\sigma\mu\nu} g_{\lambda\rho} + e_{\eta\lambda\mu\nu} g_{\sigma\rho}
	 \\&{}+ 2i \gamma_5 \bigl( g_{\lambda\rho} g_{\sigma\mu} g_{\nu\eta}
		 - g_{\sigma\rho} g_{\lambda\mu} g_{\nu\eta}
		 - g_{\lambda\rho} g_{\sigma\nu} g_{\mu\eta}
		 \\&{}+ g_{\sigma\rho} g_{\lambda\nu} g_{\mu\eta} \bigr)
	 + 2i \gamma_5 e_{\mu\nu\eta\omega} e_{\lambda\sigma\rho}{}^\omega
	 \Bigr] \mathsf{P}^{\rho[\lambda\sigma]}{}_{\bar\alpha}.
\end{split}\end{align}
This is antisymmetric in the indexes $\mu$ and $\nu$. It is also evident that each term in the above expression vanishes by virtue of the identities \eqref{P tr-less} and \eqref{eP identity}. Thus, adding the tensor matrix \eqref{odd rank nill} to the right-hand side of Eq.~\eqref{G(P)} does not affect the equality. In doing so we find that Eq.~\eqref{G(P)} takes the form
\begin{align}\label{odd-rank G 0}
	\Gamma_{[\mu\nu]\bar\alpha} = 
	\frac{i}{2} \Gamma_{[\mu\nu][\lambda\sigma]} \gamma_\rho \gamma_5 \mathsf{P}^{\rho[\lambda\sigma]}{}_{\bar\alpha}.
\end{align}
Here the matrix $\Gamma_{[\mu\nu][\lambda\sigma]}$ is defined as
\begin{equation}\label{kernel}
	\Gamma_{[\mu\nu][\lambda\sigma]}
	=
	-\frac16 \left( \sigma_{\mu\nu} \sigma_{\lambda\sigma} 
	               +3 \sigma_{\lambda\sigma} \sigma_{\mu\nu}
	         \right),
\end{equation}
It is easy to prove that the matrix $\Gamma_{[\mu\nu][\lambda\sigma]}$ satisfies the $\gamma$-tracelessness condition
\begin{align}
	\gamma^\mu \Gamma_{[\mu\nu][\lambda\sigma]} = 0 = \Gamma_{[\mu\nu][\lambda\sigma]} \gamma^\lambda
\end{align}
and is related to the projector
\begin{align}
	P_{\mu\lambda}^{(3/2)}(q)
	= \frac{1}{q^2} q^\nu q^\sigma \Gamma_{[\mu\nu][\lambda\sigma]}.
\end{align}

By making the substitution of $\Gamma_{[\mu\nu]\bar\alpha}$ for $-i\gamma_5 \Gamma_{[\mu\nu]\bar\alpha}$, we get the decomposition similar to Eq.~\eqref{odd-rank G 0}, but with the coefficients being of the same parity as $\Gamma_{[\mu\nu]\bar\alpha}$,
\begin{align}
	\Gamma_{[\mu\nu]\bar\alpha}
	= \frac12 \Gamma_{[\mu\nu][\lambda\sigma]} \gamma_\rho
	   \mathsf{S}^{\rho[\lambda\sigma]}{}_{\bar\alpha},
\end{align}
where the coefficients $\mathsf{S}_{\rho[\lambda\sigma]\bar\alpha}$ are defined by Eq.~\eqref{def of S} and satisfy the same conditions as $\mathsf{P}_{\rho[\lambda\sigma]\bar\alpha}$ do,
\begin{align}\label{constraints on S}
	\mathsf{S}^\lambda{}_{[\lambda\sigma]\bar\alpha} = 0,
	\qquad
	e^{\eta\rho\lambda\sigma} \mathsf{S}_{\rho[\lambda\sigma]\bar\alpha} = 0.
\end{align}
Therefore, the basis set of the odd-rank matrices constrained by Eq.~\eqref{eq:A1} consists of only one $\gamma$-traceless element $\Gamma_{[\mu\nu][\lambda\sigma]} \gamma_\rho $. The problem of finding all possible odd-rank $\gamma$-traceless matrices reduces to finding all the tensor coefficients $\mathsf{S}_{\rho[\lambda\sigma]\bar\alpha}$, which are antisymmetric in the indexes $\lambda$, $\sigma$ and subjected to the conditions \eqref{constraints on S}.


\subsubsection{\label{even rank}Matrices of even tensor ranks}

To construct the even-rank $\gamma$-traceless matrices, we should follow the same logic as in the case of the odd ranks, although the calculations become more cumbersome.
The most general decomposition of even-rank matrices can be written as
\begin{align}\label{even-rank decomposition}
	\Gamma_{[\mu\nu]\bar\alpha}
	 = \mathsf{R}_{[\mu\nu]\bar\alpha}
	   + \gamma_5 \mathsf{V}_{[\mu\nu]\bar\alpha}
	   + i\sigma^{\rho\omega} \mathsf{Q}_{[\rho\omega][\mu\nu]\bar\alpha},
\end{align}
where the tensor coefficients are given by
\begin{align}
	\mathsf{R}_{[\mu\nu]\bar\alpha}
	&{} = \frac14 \Tr \bigl( \Gamma_{\mu\nu\bar\alpha} \bigr),
	\\
	\mathsf{V}_{[\mu\nu]\bar\alpha}
	&{} = \frac14 \Tr \bigl( \gamma_5 \Gamma_{\mu\nu\bar\alpha} \bigr),
	\\
	\mathsf{Q}_{[\rho\omega][\mu\nu]\bar\alpha}
	&{} = \frac18 \Tr \bigl( i\sigma_{\rho\omega} \Gamma_{[\mu\nu]\bar\alpha} \bigr).
\end{align}
Imposing the requirement of $\gamma$ tracelessness on the decomposition \eqref{even-rank decomposition}, we get
\begin{align}
	\gamma^\mu \Gamma_{[\mu\nu]\bar\alpha}
	 = {}&\gamma^\mu \left[ \mathsf{R}_{[\mu\nu]\bar\alpha}
	   + i \left( \delta^\omega_\mu g^{\rho\lambda} - \delta^\rho_\mu g^{\omega\lambda} \right)
	     \mathsf{Q}_{[\rho\omega][\lambda\nu]\bar\alpha} \right]
	   \notag\\
	   &{} + i\gamma^\mu \gamma_5 \left[ -i \mathsf{V}_{[\mu\nu]\bar\alpha}
	     - i e^{\lambda\rho\omega}{}_\mu \mathsf{Q}_{[\rho\omega][\lambda\nu]\bar\alpha} \right] =0,
\end{align}
where we have used the identity
\begin{align}
	\gamma_\lambda \sigma_{\mu\nu} =
		\left( g_{\nu\eta} g_{\mu\lambda} - g_{\mu\eta} g_{\nu\lambda} \right) \gamma^\eta
		 - i e_{\mu\nu\lambda\eta} \gamma^\eta \gamma_5.
\end{align}
Since the Dirac matrices $\gamma_\mu$ and $i\gamma_\mu \gamma_5$ are mutually orthogonal, the coefficients at these matrices equal zero. Explicitly separating symmetric and antisymmetric in $\mu$ and $\nu$ parts of the coefficients, we obtain the relations
\begin{align}\label{S(Q)}
	\mathsf{R}_{[\mu\nu]\bar\alpha}
	 &{}= i \kappa''_{[\mu\nu]}{}^{[\rho\omega][\lambda\sigma]}
	      \mathsf{Q}_{[\rho\omega][\lambda\sigma]\bar\alpha},
	\\ \label{P(Q)}
	\mathsf{V}_{[\mu\nu]\bar\alpha}
	 &{}= \epsilon''_{[\mu\nu]}{}^{[\rho\omega][\lambda\sigma]}
	      \mathsf{Q}_{[\rho\omega][\lambda\sigma]\bar\alpha},
\end{align}
and two tensor constraints imposed on the coefficients $\mathsf{Q}_{[\rho\omega][\lambda\sigma]\bar\alpha}$
\begin{align}\label{kappa-identity}
	\kappa'_{(\mu\nu)}{}^{[\rho\omega][\lambda\sigma]}
	 \mathsf{Q}_{[\rho\omega][\lambda\sigma]\bar\alpha} &{}=0,
	\\ \label{epsilon-identity}
	\epsilon'_{(\mu\nu)}{}^{[\rho\omega][\lambda\sigma]}
	 \mathsf{Q}_{[\rho\omega][\lambda\sigma]\bar\alpha} &{}=0.
\end{align}
In Eqs.~\eqref{S(Q)}--\eqref{epsilon-identity} we introduce the tensors defined as
\begin{multline}\label{kappa'}
	\kappa'_{(\mu\nu)}{}^{[\rho\omega][\lambda\sigma]}
	\\
	=
	\frac14 \bigl(   \delta^\rho_\mu \delta^\sigma_\nu g^{\omega\lambda}
	               - \delta^\omega_\mu \delta^\sigma_\nu g^{\rho\lambda}
	               + \delta^\rho_\nu \delta^\sigma_\mu g^{\omega\lambda}
	               - \delta^\omega_\nu \delta^\sigma_\mu g^{\rho\lambda}
	\\{}
	               - \delta^\rho_\mu \delta^\lambda_\nu g^{\omega\sigma}
	               + \delta^\omega_\mu \delta^\lambda_\nu g^{\rho\sigma}
	               - \delta^\rho_\nu \delta^\lambda_\mu g^{\omega\sigma}
	               + \delta^\omega_\nu \delta^\lambda_\mu g^{\rho\sigma}
	        \bigr),
\end{multline}
\begin{multline}
	\kappa''_{[\mu\nu]}{}^{[\rho\omega][\lambda\sigma]}
	\\{}
	=
	\frac14 \bigl(   \delta^\rho_\mu \delta^\sigma_\nu g^{\omega\lambda}
	               - \delta^\omega_\mu \delta^\sigma_\nu g^{\rho\lambda}
	               - \delta^\rho_\nu \delta^\sigma_\mu g^{\omega\lambda}
	               + \delta^\omega_\nu \delta^\sigma_\mu g^{\rho\lambda}
	\\{}
	               - \delta^\rho_\mu \delta^\lambda_\nu g^{\omega\sigma}
	               + \delta^\omega_\mu \delta^\lambda_\nu g^{\rho\sigma}
	               + \delta^\rho_\nu \delta^\lambda_\mu g^{\omega\sigma}
	               - \delta^\omega_\nu \delta^\lambda_\mu g^{\rho\sigma}
	        \bigr),
\end{multline}
\begin{multline}\label{epsilon'}
	\epsilon'_{(\mu\nu)}{}^{[\rho\omega][\lambda\sigma]}
	\\
	=
	\frac14 \bigl( - e^{\lambda\rho\omega}{}^{\vphantom{\mu}}_\nu \delta^\sigma_\mu
	               - e^{\lambda\rho\omega}{}^{\vphantom{\mu}}_\mu \delta^\sigma_\nu
	               + e^{\sigma\rho\omega}{}^{\vphantom{\mu}}_\nu \delta^\lambda_\mu
	               + e^{\sigma\rho\omega}{}^{\vphantom{\mu}}_\mu \delta^\lambda_\nu
	        \bigr),
\end{multline}
\begin{multline}
	\epsilon''_{[\mu\nu]}{}^{[\rho\omega][\lambda\sigma]}
	\\
	=
	\frac14 \bigl(   e^{\lambda\rho\omega}{}^{\vphantom{\mu}}_\nu \delta^\sigma_\mu
	               - e^{\lambda\rho\omega}{}^{\vphantom{\mu}}_\mu \delta^\sigma_\nu
	               - e^{\sigma\rho\omega}{}^{\vphantom{\mu}}_\nu \delta^\lambda_\mu
	               + e^{\sigma\rho\omega}{}^{\vphantom{\mu}}_\mu \delta^\lambda_\nu
	        \bigr).
\end{multline}
In the above equations square brackets emphasize antisymmetric pairs of indices and round ones denote symmetric pairs.

Substituting Eqs.~\eqref{S(Q)} and \eqref{P(Q)} into the decomposition \eqref{even-rank decomposition}, we get
\begin{multline}\label{even-rank G 0}
	\Gamma_{\vphantom{[}[\mu\nu]\bar\alpha}^{\vphantom{}}
	 = \biggl[ i \kappa''_{[\mu\nu][\rho\omega][\lambda\sigma]}
	          + \gamma_5 \epsilon''_{[\mu\nu][\rho\omega][\lambda\sigma]}
	          \\{}
	          + \frac{i}2 \sigma_{\rho\omega}
	            \bigl( g_{\mu\lambda} g_{\nu\sigma} - g_{\nu\lambda} g_{\mu\sigma} \bigr)
	   \biggr] \mathsf{Q}^{[\rho\omega][\lambda\sigma]}{}_{\bar\alpha}.
\end{multline}
Here the matrix in the square brackets can be made manifestly $\gamma$ traceless by adding the following expression to the right-hand side of Eq.~\eqref{even-rank G 0} 
\begin{widetext}
\begin{align}
	-\frac{i}2 \biggl[
	   \left( \kappa'_{(\gamma\zeta)[\mu\nu][\eta\xi]}
	      +\frac16 \left( g_{\mu\eta} g_{\nu\xi} - g_{\mu\xi} g_{\nu\eta} \right) g_{\gamma\zeta}
	   \right) \kappa'^{(\gamma\zeta)}{}_{[\rho\omega][\lambda\sigma]}
	  +\left( \epsilon'_{(\gamma\zeta)[\mu\nu][\eta\xi]}
	      +\frac13 e_{\mu\nu\eta\xi} g_{\gamma\zeta}
	   \right) \epsilon'^{(\gamma\zeta)}{}_{[\rho\omega][\lambda\sigma]}
	\biggr] \sigma^{\eta\xi} \mathsf{Q}^{[\rho\omega][\lambda\sigma]}{}_{\bar\alpha},
\end{align}
which is equal to zero owing to the identities \eqref{kappa-identity} and \eqref{epsilon-identity}. Finally, the most general decomposition of even-rank matrix satisfying Eq.~\eqref{eq:A1} takes the form
\begin{align}\label{even-rank G 1}
	\Gamma_{[\mu\nu]\bar\alpha}
	 = -\frac{3i}{8} \left[ \left(
	     \Gamma_{[\mu\nu][\rho\lambda]} g_{\omega\sigma}
	    -\Gamma_{[\mu\nu][\omega\lambda]} g_{\rho\sigma}
	    -\Gamma_{[\mu\nu][\rho\sigma]} g_{\omega\lambda}
	    +\Gamma_{[\mu\nu][\omega\sigma]} g_{\rho\lambda} \right)
	    -\frac13 \left( \Gamma_{[\mu\nu][\rho\omega]} \sigma_{\lambda\sigma}
	                   +\Gamma_{[\mu\nu][\lambda\sigma]} \sigma_{\rho\omega} \right)
	   \right] \mathsf{Q}^{[\rho\omega][\lambda\sigma]}{}_{\bar\alpha},
\end{align}
\end{widetext}
where $\Gamma_{[\mu\nu][\lambda\sigma]}$ is $\gamma$-traceless matrix introduced by Eq.~\eqref{kernel}.

Note that the first term in square brackets in Eq.~\eqref{even-rank G 1} is antisymmetric under interchange of the pairs of indexes $[\lambda\sigma]$ and $[\rho\omega]$, while the second one is symmetric. We have thereby come to the conclusion that there are two independent even-rank basis matrices constrained by Eq.~\eqref{eq:A1}. To express this manifestly, we can rewrite Eq.~\eqref{even-rank G 1} as
\begin{align}\label{even-rank G 2}
	\Gamma_{[\mu\nu]\bar\alpha}
	 = \Gamma_{[\mu\nu][\lambda\sigma]}
	   \mathsf{R}^{[\lambda\sigma]}{}_{\bar\alpha}
	  +\frac{i}{2} \Gamma_{[\mu\nu][\lambda\sigma]} \sigma_{\rho\omega}
	   \mathsf{Q}^{[\rho\omega][\lambda\sigma]}{}_{\bar\alpha}.
\end{align}
Here we have used Eq.~\eqref{S(Q)} and the following identity:
\begin{align}
		& \Gamma_{[\mu\nu][\rho\omega]} \sigma_{[\lambda\sigma]}
	-\Gamma_{[\mu\nu][\lambda\sigma]} \sigma_{[\rho\omega]}
  +\Gamma_{[\mu\nu][\lambda\rho]} g_{\sigma\omega}
\notag\\&
	-\Gamma_{[\mu\nu][\sigma\rho]} g_{\lambda\omega}
	-\Gamma_{[\mu\nu][\lambda\omega]} g_{\sigma\rho}
	+\Gamma_{[\mu\nu][\sigma\omega]} g_{\lambda\rho}
	= 0.
\end{align}


\subsection{\label{app:a2}$\gamma$-traceless matrices for $\ell \geqslant 2$}

The general matrix $\Gamma_{\bar A \bar\alpha}$ for $\ell \geqslant 2$ can be written as follows:
\begin{align}\label{eq:A39}
	\Gamma_{\bar A \bar\alpha}
	={}&
	\mathsf{R}_{\bar A \bar\alpha}
	+ i \gamma_5 \mathsf{V}_{\bar{A} \bar\alpha}
	+ i \sigma^{C} \mathsf{Q}_{C\bar{A}\bar\alpha}
	\notag\\&{}
	+\gamma^\eta \mathsf{S}_{\eta\bar{A}\bar\alpha}
	+ i \gamma^\eta \gamma_5 \mathsf{P}_{\eta\bar{A}\bar\alpha},
\end{align}
where tensor coefficients are given by
\begin{align}\label{eq:R}
	\mathsf{R}_{\bar A \bar\alpha}
	&{} = \frac14 \Tr \bigl( \Gamma_{\bar A \bar\alpha} \bigr),
	\\
	\mathsf{V}_{\bar A \bar\alpha}
	&{} = \frac14 \Tr \bigl( \gamma_5 \Gamma_{\bar A \bar\alpha} \bigr),
	\\
	\mathsf{Q}_{C\bar A \bar\alpha}
	&{} = \frac18 \Tr \bigl( i\sigma_C \Gamma_{\bar A \bar\alpha} \bigr),
	\\ \label{eq:S}
	\mathsf{S}_{\eta\bar A \bar\alpha}
	&{} = \frac14 \Tr \bigl( \gamma_\eta \Gamma_{\bar A \bar\alpha} \bigr),
	\\ \label{eq:P}
	\mathsf{P}_{\eta\bar A \bar\alpha}
	&{} = \frac14 \Tr \bigl( i\gamma_\eta \gamma_5 \Gamma_{\bar A \bar\alpha} \bigr).
\end{align}
Here and in what follows multi-indices are defined as
\begin{equation}\label{app:indices}
\begin{gathered}
	A^a = [\mu_a\nu_a], \qquad \bar A = (A^1 \dots A^\ell),
	\\
	\bar A^a = (A^1 \dots A^{a-1} A^{a+1} \dots A^\ell),
	\\
	B^b = [\lambda_b\sigma_b], \qquad \bar B = (B^1 \dots B^\ell),
	\\
	\bar B^b = (B^1 \dots B^{b-1} B^{b+1} \dots B^\ell),
	\\
	C = [\rho\omega].
\end{gathered}
\end{equation}
The tensor coefficients are subjected to a number of constraints following from the condition \eqref{eq:A1}:
\begin{alignat}{2}\label{eq:A46}
	g^{\mu^i\mu^j} \mathsf{H} &= 0, &
	e^{A^i A^j} \mathsf{H} &= 0,
	\\ \label{eq:A47}
	e^{A^i \mu^j \mu^k} \mathsf{H} &= 0, &
	e^{\mu^i \mu^j \mu^k \mu^l} \mathsf{H} &= 0,
	\\ \label{eq:A48}
	e^{CA^i} \mathsf{Q}_{C\bar A \bar\alpha} &= 0,&\qquad
	g^{\rho\mu^i} g^{\omega\mu^j} \mathsf{Q}_{C\bar A \bar\alpha} &= 0,
\end{alignat}
where $i$, $j$, $k$, $l = 1,\,2,\dots\ell$ and $\mathsf{H}$ is any of the tensor coefficients \eqref{eq:R}--\eqref{eq:P}. The relations \eqref{eq:A48} follow directly from Eqs.~\eqref{kappa-identity} and \eqref{epsilon-identity}. The relations \eqref{eq:A46} and \eqref{eq:A47} are easily proved by well-known identities for products of the Dirac matrices $\gamma_\mu \gamma_\nu$, $\gamma_\mu \gamma_\nu \gamma_\lambda \gamma_\sigma$. In particular,
\begin{align}\label{eq:A49}
	0 = \gamma^{\mu^i} \gamma^{\mu^j} \Gamma_{\bar A \bar\alpha}
	= g^{\mu^i\mu^j} \Gamma_{\bar A \bar\alpha}.
\end{align}
The first of the relations \eqref{eq:A46} follows from Eq.~\eqref{eq:A49} and the definitions \eqref{eq:R}--\eqref{eq:P}. Other relations in Eqs.~\eqref{eq:A46} and \eqref{eq:A47} can be proved similarly.

A set of constraints for the coefficients $\mathsf{S}_{\eta\bar A \bar\alpha}$ and $\mathsf{P}_{\eta\bar A \bar\alpha}$ follows from Eqs.~\eqref{S tr-less}, \eqref{eP identity}, and \eqref{constraints on S}:
\begin{alignat}{2}\label{eq:A50}
	e^{\rho\eta A^i} \mathsf{S}_{\eta\bar A \bar\alpha} &= 0, &\qquad
	g^{\rho\mu^i } \mathsf{S}_{\eta\bar A \bar\alpha} &= 0,
	\\ \label{eq:A51}
	e^{\rho\eta A^i} \mathsf{P}_{\eta\bar A \bar\alpha} &= 0, &\qquad
	g^{\rho\mu^i } \mathsf{P}_{\eta\bar A \bar\alpha} &= 0
\end{alignat}
for arbitrary $i = 1,\,2 \dots \ell$.

The coefficients $\mathsf{Q}_{C\bar A \bar\alpha}$ can be decomposed into two parts:
\begin{align}
	\mathsf{Q}_{C\bar A \bar\alpha}
	&= \mathsf{T}_{C\bar A \bar\alpha} +\mathsf{U}_{C\bar A \bar\alpha},
	\\ \label{eq:A53}
	\mathsf{T}_{C\bar A \bar\alpha}
	&= \frac12 \Bigl( \mathsf{Q}_{C\bar A \bar\alpha}
	                 + \sum_{a=1}^\ell \mathsf{Q}_{A_a C \bar A^a \bar\alpha}
	           \Bigr),
	\\ \label{eq:A54}
	\mathsf{U}_{C\bar A \bar\alpha}
	&= \frac12 \Bigl( \mathsf{Q}_{C\bar A \bar\alpha}
	                 - \sum_{a=1}^\ell \mathsf{Q}_{A_a C \bar A^a \bar\alpha}
	           \Bigr).
\end{align}
For the tensors \eqref{eq:A53} and \eqref{eq:A54} the relations \eqref{eq:A46}--\eqref{eq:A48} give the following:
\begin{alignat}{2}
	e^{CA^i} \mathsf{U}_{C\bar A \bar\alpha} &= 0, &\qquad
	e^{CA^i} \mathsf{T}_{C\bar A \bar\alpha} &= 0,
	\\ \label{eq:A56}
	e^{A^i A^j} \mathsf{U}_{C\bar A \bar\alpha} &= 0, &\qquad
	e^{A^i A^j} \mathsf{T}_{C\bar A \bar\alpha} &= 0,
	\\ \label{eq:A57}
	g^{\mu^i\mu^j} \mathsf{T}_{C\bar A \bar\alpha} &= 0,&
	g^{\rho\mu^i} \mathsf{T}_{C\bar A \bar\alpha} &= 0,
	\\
	g^{\mu^i\mu^j} \mathsf{U}_{C\bar A \bar\alpha} &= 0,&\qquad
	g^{\rho\mu^i} g^{\rho\mu^j} \mathsf{U}_{C\bar A \bar\alpha} &= 0.
\end{alignat}

Now we can prove that any traceless matrix $\Gamma_{\bar A\bar\alpha}$ for any $\ell$ can be decomposed as follows,
\begin{align}\label{app:total decomposition}
	\Gamma_{\bar A\bar\alpha}
	 = \frac{1}{4^{\ell-1}} \Gamma_{\bar A \bar B}
	   \left[ \mathsf{R}^{\bar B}{}_{\bar\alpha}
	         + \frac12 \gamma_\rho \mathsf{S}^{\rho\bar B}{}_{\bar\alpha}
	         + \frac{i}2 \sigma_{\rho\omega} \mathsf{Q}^{[\rho\omega]\bar B}{}_{\bar\alpha}
	   \right],
\end{align}
if the decomposition \eqref{app:total decomposition} is valid for $\ell-1$. The matrix $\Gamma_{\bar A \bar B}$ for arbitrary $\ell$ is defined recursively:
\begin{align}\label{eq:A60}
	\Gamma_{\bar A \bar B}^{(\ell)}
	&{} = \frac{3}{2(2\ell+1)\ell^2}\sum_{a,b=1}^\ell \biggl[ (\ell+1) \Gamma^{(\ell-1)}_{\bar A^a \bar B^b} \Gamma^{(1)}_{A_a B_b}
	+{}
		\notag\\
	&{}+(\ell-1) \Gamma^{(\ell-1)}_{\bar A^a \bar B^b} \Gamma^{(1)}_{B_b A_a}
	- \sum_{b \ne c=1}^{\ell} \Gamma^{(\ell-1)}_{\bar A^a A^a\bar B^{bc}}
	\Gamma^{(1)}_{B_b B_c} \biggr].
\end{align}
From now on summation symbols are omitted for brevity, but summation over repeated lower-case Latin indices is implied, if one of the indices is in a subscript position and the other is in a superscript one (e.g. $\bar{A}^a A_a$).

If the decomposition \eqref{app:total decomposition} is valid for $\ell-1$, then the decomposition \eqref{eq:A39} for $\ell$ can be rewritten as
\begin{align}\label{eq:A62}
	\Gamma_{\bar A\bar\alpha}
	 = {}&\frac{1}{4^{\ell-2}} \frac{1}{2\ell^2}
	   \Gamma_{\bar A^a \bar B^b} g_{A_a B_b}
	\notag\\& \times
	   \left[ \mathsf{R}^{\bar B}{}_{\bar\alpha}
	         + \frac12 \gamma_\rho \mathsf{S}^{\rho\bar B}{}_{\bar\alpha}
	         + \frac{i}2 \sigma_{C} \mathsf{Q}^{C\bar B}{}_{\bar\alpha}
	   \right],
\end{align}
where
\begin{align}
	g_{A_a B_b} = g_{[\mu_a\nu_a][\lambda_b\sigma_b]}
	= g_{\mu_a\lambda_b} g_{\nu_a\sigma_b} - g_{\mu_a\sigma_b} g_{\nu_a\lambda_b}.
\end{align}
Let us consider the term with the coefficient $\mathsf{T}^{C\bar B}{}_{\bar\alpha}$:
\begin{multline}\label{eq:A64}
	\frac{1}{4^{\ell-2}} \frac{1}{2\ell^2}
	\Gamma_{\bar A^a \bar B^b} g_{A_a B_b}
	\sigma_{C} \mathsf{T}^{C\bar B}{}_{\bar\alpha}
	\\=
	\frac{1}{4^{\ell-1}} \frac{1}{\ell^2}
	\Gamma_{\bar A^a \bar B^b} (g_{A_a B_b} - i\gamma_5 e_{A_a B_b})
	\sigma_{C} \mathsf{T}^{C\bar B}{}_{\bar\alpha}
	\\=
	\frac{1}{4^{\ell-1}} \frac{1}{2\ell^2}
	\Gamma_{\bar A^a \bar B^b} [\sigma_{A_a},\, \sigma_{B_b} ]_+
	\sigma_{C} \mathsf{T}^{C\bar B}{}_{\bar\alpha},
\end{multline}
where
\begin{align}
	[\sigma_{A_a},\, \sigma_{B_b} ]_+
	&= \sigma_{A_a} \sigma_{B_b} + \sigma_{B_b} \sigma_{A_a}
	\notag\\
	&= -2 (g_{A_a B_b} - i\gamma_5 e_{A_a B_b}).
\end{align}
The equation \eqref{eq:A64} is derived using the identities \eqref{eq:A57} and a relation
\begin{align}
	i\gamma_5 \Gamma_{\bar A \bar B} = \frac12 e^{B^i C} \Gamma_{\bar A \bar B^i C}
\end{align}
that is an immediate consequence of Eq.~\eqref{eq:A6} and the recursive definition of $\Gamma_{\bar A \bar B}$ \eqref{kernel} and \eqref{eq:A60}.

Due to the properties \eqref{eq:A56} and \eqref{eq:A57} of the coefficients $\mathsf{T}_{C\bar B\bar\alpha}$, we have $\sigma_{B_b} \sigma_{C} \mathsf{T}^{C\bar B}{}_{\bar\alpha} = 0 = \sigma_{B_a} \sigma_{B_b} \mathsf{T}^{C\bar B}{}_{\bar\alpha}$. Hence, we can recast \eqref{eq:A64} as
\begin{multline}\label{eq:A68a}
	\frac{1}{4^{\ell-2}} \frac{1}{2\ell^2}
	\Gamma_{\bar A^a \bar B^b} g_{A_a B_b}
	\sigma_{C} \mathsf{T}^{C\bar B}{}_{\bar\alpha}
	=
	\frac{1}{4^{\ell-1}} \Gamma_{\bar A \bar B}
	\sigma_{C} \mathsf{T}^{C\bar B}{}_{\bar\alpha}.
\end{multline}
To complete the proof of Eq.~\eqref{app:total decomposition}, the terms with $\mathsf{S}^{\rho\bar B\alpha}$ and $\mathsf{R}^{\bar B\alpha} + \frac{i}2 \sigma_{C} \mathsf{U}^{C\bar B}{}_{\bar\alpha}$ in the decomposition \eqref{eq:A62} can be transformed in the same way as Eqs.~\eqref{eq:A62}--\eqref{eq:A68a}. As a result, we come to a conclusion that any $\gamma$-traceless matrix $\Gamma_{\bar{A}\bar\alpha}$ for arbitrarily high $\ell$ can be written as the decomposition \eqref{app:total decomposition}.


\subsection{\label{app:a3}Algebraic properties of the $\gamma$-traceless basis matrices}

The martix \eqref{eq:A60} is self-conjugate:
\begin{align}\label{eq:A68}
	\bar\Gamma_{\bar A \bar B} = \Gamma_{\bar B \bar A}.
\end{align}
This can be checked straightforwardly for $\ell = 1$ and 2. For higher $\ell \geqslant 2$ the property \eqref{eq:A68} can be proved by using the recursive definition \eqref{eq:A60} twice. We find that if Eq.~\eqref{eq:A68} holds for $\ell-1$ and $\ell-2$, then it holds for $\ell$.

The $\gamma$ tacelessness of the matrix $\Gamma_{\bar A \bar B}$,
\begin{align}\label{eq:A69}
	\gamma^{\mu_1} \Gamma_{\bar A \bar B} = 0 = \Gamma_{\bar A \bar B} \gamma^{\lambda_1},
\end{align}
can now be also proved. The right-hand side of Eq.~\eqref{eq:A69} is proved by using Eqs.~\eqref{kernel} and \eqref{eq:A60} and the following relations that are implied to hold for $\ell-1$:
\begin{align}\label{eq:A70}
	 \Gamma_{\bar{A}\bar{B}^\ell[\lambda_\ell\sigma_\ell]} \gamma_\rho
	+\Gamma_{\bar{A}\bar{B}^\ell[\rho\lambda_\ell]} \gamma_{\sigma_\ell}
	+\Gamma_{\bar{A}\bar{B}^\ell[\sigma_\ell\rho]} \gamma_{\lambda_\ell} = 0,
\end{align}
\begin{align}\label{eq:A71}
	& \Gamma_{\bar{A}\bar{B}^\ell[\rho\omega]} \sigma_{[\lambda_\ell\sigma_\ell]}
	-\Gamma_{\bar{A}\bar{B}^\ell[\lambda_\ell\sigma_\ell]} \sigma_{[\rho\omega]}
  +\Gamma_{\bar{A}\bar{B}^\ell[\lambda_\ell\rho]} g_{\sigma_\ell\omega}
\notag\\&
	-\Gamma_{\bar{A}\bar{B}^\ell[\sigma_\ell\rho]} g_{\lambda_\ell\omega}
	-\Gamma_{\bar{A}\bar{B}^\ell[\lambda_\ell\omega]} g_{\sigma_\ell\rho}
	+\Gamma_{\bar{A}\bar{B}^\ell[\sigma_\ell\omega]} g_{\lambda_\ell\rho}
	= 0.
\end{align}
The left-hand side of Eq.~\eqref{eq:A69} follows directly from the right-hand one and Eq.~\eqref{eq:A68}. Then the identity can be proved for $\ell$. Since the $\gamma$ tracelessness of the matrix $\Gamma_{\bar A \bar B}$ has been proved for $\ell$, the constraints \eqref{eq:A50} and \eqref{eq:A51} are valid for the tensors \eqref{eq:S} and \eqref{eq:P}, where $\Gamma_{\bar A \bar\alpha}$ is taken to be $\Gamma_{\bar A \bar B} \gamma_\zeta$. We have
\begin{multline}\label{eq:A72}
	e_{\omega\xi\alpha\beta} e^{\omega\eta A^i} \mathsf{S}_{\eta\bar A \bar B \zeta}
	\\=
	-2 \left( \mathsf{S}_{\xi[\alpha\beta]\bar A^i \bar B \zeta}
	         +\mathsf{S}_{\beta[\xi\alpha]\bar A^i \bar B \zeta}
	         +\mathsf{S}_{\alpha[\beta\xi]\bar A^i \bar B \zeta}
	   \right) = 0
\end{multline}
and a similar relation for $\mathsf{P}_{\eta\bar A \bar B \zeta}$. Granting the definitions \eqref{eq:S} and \eqref{eq:P} of the tensors $\mathsf{S}_{\eta\bar A \bar B \zeta}$ and $\mathsf{P}_{\eta\bar A \bar B \zeta}$, the identity \eqref{eq:A70} follows from the relation \eqref{eq:A72}.

The second identity \eqref{eq:A71} is a direct consequence of the one just proved. To show this, we have to expand matrices $\sigma_{\mu\nu}$ in Eq.~\eqref{eq:A71} as $\frac12 (\gamma_\mu \gamma_\nu - \gamma_\nu \gamma_\mu)$ and transform each term thus obtained by means of the identity~\eqref{eq:A70}.

Also it can be proved that the matrix $\Gamma_{\bar A \bar B}$ is normalized so that to give the projector for the spin-$\bigl( \ell + \frac12 \bigr)$:
\begin{align}\label{eq:A73}
	\Gamma_{\bar A \bar B} \cdot \prod_{i=1}^\ell \frac{q^{\nu_i} q^{\sigma_i}}{q^2}
	= P^{(\ell + \frac12)}_{\bar\mu\bar\lambda}(q),
\end{align}
where $\bar{\mu}=\mu_1\mu_2...\mu_\ell$, $\bar{\lambda}=\lambda_1\lambda_2...\lambda_\ell$. Indeed, since there is just one unique projector operator $P^{(\ell + \frac12)}_{\bar\mu\bar\lambda}(q)$ for any $\ell$ and the $\gamma$ tracelessness and other properties of the left-hand side of Eq.~\eqref{eq:A73} are those of the projector, then the left-hand side of Eq.~\eqref{eq:A73} is obviously proportional to the projector. The only thing we need to check is the normalization $\Tr P^{(\ell + \frac12)}_{\bar\mu}{}^{\bar\mu}(q) = 4 (\ell+1)$ \cite{1957PhRv..106..345B, springerlink:10.1007/BF02747684}. This can be done easily by induction.

Finally, using Eqs.~\eqref{kernel}, \eqref{eq:A60}, and \eqref{eq:A73}, the validity of the following identity can be checked:
\begin{multline}
	\Gamma_{\bar A\bar B} 
	P^{\bar\sigma\bar\alpha}_{(\ell+\frac12)}(p) \cdot
	\prod_{a=1}^\ell p^{\lambda_a} q^{\nu_a}
	\cdot \prod_{b=2}^\ell p^{\mu_b}
	\\=
	(p^2)^{\ell-1} p^\lambda q^\nu \Gamma_{[\mu_1\nu][\lambda\sigma_1]}
	P^{\bar\sigma\bar\alpha}_{(\ell+\frac12)}(p) \cdot
	\prod_{a=2}^\ell q_{\sigma_a}.
\end{multline}


%


\end{document}